\documentclass[a4paper,11pt]{article}

\usepackage{bm}
\usepackage{dcolumn}
\usepackage{enumerate}
\usepackage{multirow}
\usepackage{inputenc}
\usepackage{fontenc}
\usepackage{slashbox}
\usepackage{float}
\usepackage{graphicx, rotating, threeparttable}
\usepackage{geometry}
\usepackage{array, multirow}
\usepackage[export]{adjustbox}
\usepackage{subfig}
\usepackage{comment}

\pdfoutput=1 

\usepackage{jcappub} 

\usepackage[T1]{fontenc} 

\title{Current constraints on deviations from General Relativity using binning in redshift and scale}

\author[a]{Cristhian Garcia-Quintero,}
\author[a]{Mustapha Ishak}
\author[a, b]{and Orion Ning}

\affiliation[a]{Department of Physics, The University of Texas at Dallas, Richardson, Texas 75080, USA}
\affiliation[b]{Department of Physics, University of California, Berkeley, CA 94720, USA}

\emailAdd{gqcristhian@utdallas.edu}
\emailAdd{mishak@utdallas.edu}
\emailAdd{orion.ning@berkeley.edu}

\abstract
{
We constrain deviations from general relativity (GR) including \textit{both} redshift and scale dependencies in the modified gravity (MG) parameters. In particular, we employ the under-used binning approach and compare the results to functional forms. We use available datasets such as Cosmic Microwave Background (CMB) from Planck 2018,  Baryonic Acoustic Oscillations (BAO) and Redshift Space Distortions (BAO/RSD) from the BOSS Data Release 12, the 6DF Galaxy Survey, the SDSS Data Release 7 Main Galaxy Sample, the correlation of Lyman-$\alpha$ forest absorption and quasars from SDSS-DR14, Supernova Type Ia (SNe) from the Pantheon compilation, and DES Y1 data. Moreover, in order to maximize the constraining power from available datasets, we analyze MG models where we alternatively set some of the MG parameters to their GR values and vary the others. Using functional forms, we find an up to 3.5-$\sigma$ tension with GR in $\Sigma$ (while $\mu$ is fixed) when using Planck+SNe+BAO+BAO/RSD; this goes away when lensing data is included, i.e. CMB lensing and DES Y1 (CMBL+DES). Using different binning methods, we find that a tension with respect to GR above 2-$\sigma$ in the (high-z, high-k) bin is persistent even when including CMBL+DES to Planck+SNe+BAO+BAO/RSD. Also, we find another tension above 2-$\sigma$ in the (low-z, high-k) bin, but that can be reduced with the addition of lensing data. Furthermore, we perform a model comparison using the Deviance Information Criterion statistical tool and find that the MG model ($\mu=1$, $\Sigma$) is weakly favored by the data compared to $\Lambda$CDM, except when DES data is included. Another noteworthy result is that we find that the binning methods \textit{do not agree} with the widely-used functional parameterization where the MG parameters are proportional to $\Omega_{\text{DE}}(a)$, and this is clearly apparent in the high-z and high-k regime where this parameterization underestimates the deviations from GR.
}

\begin{document}
\maketitle
\flushbottom

\section{Introduction
\label{sec:intro}}
The complex problem of cosmic acceleration added a pressing motivation to testing gravity physics at cosmological scales, see, e.g., the reviews \cite{2012-Clifton-MG,2015-rev-Joyce-et-al,KOYAMA2016TestGR,2016-Joyce-Lombriser-Schmidt-DEvsMG,IshakMG2019} and references therein. Indeed, the recent years witnessed a growing and active interest in testing deviations from General Relativity (GR) using cosmological datasets, see, e.g., the partial list  \cite{2006-Ishak-splitting, GongCurved2009,   DossettFOM2011, Majerotto2012, DossettMGC2012, Lombriser2014, Hellwing2014, Ade:2015rim, JoudakiEtAl2017, DossettDEspeed, Song:2015oza, Lin2016, Burrage2016, PogosianEtAl2016,  Dossett2015, Alonso:2016suf, Linder2017OmegaDE1,  Nesseris:2017vor, BeanTangmatitham2010, Burrage2018,DESMG2018, Singh2018, Abbott:2018xao, SimpsonEtAl2013, Heymans:2018khp, SpurioMancini:2019rxy, Park:2019emi, Zucca:2019xhg, Kim:2019kls,Ferte2017,2019ApJ...871..196L,2019MG-Kazantzidis,2020MG-Skara,2018MG-Gannouji,Lensing.Anomaly2,2020MG-Raveri, Blake:2020mzy}. 

A chief approach in doing so has been the use of phenomenological parameterizations to model departures from GR at the level of the perturbed Einstein's field equations. The aim is to constrain such modified gravity (MG) parameters and determine, with some significance, whether they deviate or not from their expected GR values. 
However, some questions have been raised, for example in \cite{Linder2017,Lagos-RunningPlanckMass,LinderEtAl2015}, about the limitation of some specific functional forms for these parameters which can fail in capturing a deviation from GR \cite{Linder2017}.  

In this paper, we use a binning approach as an alternative to functional forms and then we compare results from the two approaches. We also include both redshift and scale dependencies for MG parameters. It was stressed in \cite{DossettDEspeed,PogosianEtAl2016} that scale dependence can play a unique role in distinguishing between dark energy and modified gravity. We use a reconstruction approach to perform a comparison between functional and binning methods which include various bins in scale and redshift. We also choose a strategy of varying only a subset of modified gravity parameters at a time in order to maximize the constraining power provided by current available data sets. We use 
temperature and polarization data, as well as lensing from Planck 2018 \cite{Planck2018,Planck-2018-lensing},
 Baryonic Acoustic Oscillations (BAO), Redshift Space Distortions (BAO/RSD) from the BOSS Data Release 12 \cite{AlamEtAl2016}, 
  the 6DF Galaxy Survey \cite{BeutlerEtAl2011}, 
  the SDSS Data Release 7 Main Galaxy Sample \cite{RossEtAl2014}, 
  BAO from the correlation of Lyman-$\alpha$ forest absorption 
  and quasars from SDSS-DR14 \cite{2019-BAO-lyalpha-quasar-Blomqvist}, 
  Supernova from the Pantheon compilation \cite{Pan-STARRS-2017}, 
  and DES Y1 clustering and lensing data \cite{DES2017}. 
 We also perform a model comparison between MG and $\Lambda$CDM using the Deviance Information Criterion statistical tool \cite{DIC-Spiegelhalter,DIC-Spiegelhalter2}.

The paper is organized as follows: After the introduction in section \ref{sec:intro}, we introduce the modified gravity (MG) parameter formalism for functional and binning methods in sections \ref{sec:MG-equations} and \ref{sec:parameterizations}. The datasets used are described in section \ref{sec:datasets}. We present and discuss the results in section \ref{sec:results}. A summary and concluding remarks are given in section \ref{sec:conclusion}.

\section{Modified gravity (MG) parameter formalism \label{sec:MG-equations}}
As stated earlier, the perturbed Einstein equations are modified by adding phenomenological parameters. The purpose of these parameters is to model  departures from GR when analyzing data from different cosmological probes. An adequate framework to apply this formalism is to use the so-called conformal Newtonian gauge, where the flat Friedmann-Lema\^{i}tre-Robertson-Walker (FLRW) metric with scalar perturbations can be written as
\begin{equation}
ds^2=a(\tau)^2[-(1+2\Psi)d\tau^2+(1-2\Phi)\delta_{ij}dx^i dx^j].
\label{eq:line-element}
\end{equation}
A further treatment based on perturbations leads to the evolution equations of the gravitational potentials $\Psi$ and $\Phi$. In Fourier space, the evolution of $\Phi$ is described by a relativistic Poisson equation while a second equation relates the two potentials which, in general, are functions of time and scale. In the late time universe, it is expected that the anisotropic stress is negligible so that the two potentials are approximately equal. Then, significant observations of a deviation from this equality should be an indication of departure from GR. This leads to defining a gravitational slip parameter as
\begin{equation}
\eta(a,k) \approx \frac{\Phi}{\Psi}.
\label{eq:definition_eta}
\end{equation}
A second MG parameter of interest is one that enters in the coupling between the gravitational potential and the source content, and which allows measuring deviations from the growth of structure as predicted by GR. Such a parameter can be defined by
\begin{equation}
k^2\Psi = -4\pi G a^2\mu(a,k) \sum_i\rho_i\Delta_i,
\label{eq:definition_mu}
\end{equation}
where $\mu(a,k)$ measures the strength of the gravitational interaction. Since this parameter can be written as $\mu=G_{\text{eff}}/G$, it can be interpreted as modification of the Newton constant of gravitation. Additionally, due to galaxy weak lensing measurements by ongoing and past surveys, it is convenient to consider an MG parameter that models departures from GR when studying gravitational lensing phenomena and the propagation of the null rays. We define $\Sigma(a,k)$ as the MG parameter that quantifies the response of massless particles in a gravitational field and it is given by
\begin{equation}
k^2(\Phi+\Psi)=-8\pi G a^2 \Sigma (a,k) \sum_i\rho_i\Delta_i,
\label{eq:definition_Sigma}
\end{equation}
where we can notice that $\Sigma$ is directly related to the so-called Weyl potential, $(\Phi+\Psi)/2$. It is worth noting that in the late time universe these MG parameters are simply related by $\Sigma=\frac{\mu}{2}(\eta+1)$. Also, we recover GR when each MG parameter is equal to 1. Hence, significant and persistent deviations of these MG parameters from their GR values would imply either deviations from the predictions of our current theory or some experimental systematic effect.

\section{MG parameterizations \label{sec:parameterizations}}
One of the important points for the problem of constraining modified gravity is to find a suitable parameterization for the MG parameters that can capture essential physical aspects to efficiently match theory and observations. As we mentioned earlier, this is not an easy task and some of the currently often-used parameterizations have not been clear from criticism, see, e.g., \cite{Linder2017,Lagos-RunningPlanckMass,LinderEtAl2015}. A plausible strategy that may shed some light in this complicated task is the use of bins for the MG parameters in both redshift and scale. Tackling the problem in this way avoids assuming an appropriated functional form, which can be challenging to formulate since it has to cover all redshift ranges and can involve scale dependence. However, even though the binning methods are the main purpose of this analysis, we also define a common functional form parameterization in order to compare both methods and, in particular, the scale dependence, which has not been thoroughly studied even in the functional form cases. In the following, we define the parameterizations used in this work.

\subsection{Functional form \label{sec:functional}}
Very often, the MG parameters are assumed to only have time dependence. A commonly used ansatz for the MG parameters is to assume a time dependence based on the $\Omega_{\text{DE}}$ parameter. However, it has been shown that in near- and super-horizon scales we can consider a specific scale dependence for the MG parameters as ratios of polynomials in $k$ \cite{BakerEtAl2014b,2013PhRvD..87j4015S}. Therefore, we use a scale-dependent factor that adds three extra parameters. We base our analysis on $\mu$ and $\Sigma$, which are parameters directly constrained by the cosmological probes (as opposed to $\eta$, which is a derived parameter) and model them as
\begin{equation}
\mu(a,k)=1+\mu_{0}\frac{\Omega_{\text{DE}}(a)}{\Omega_\Lambda} \left[ \frac{1+c_1 \left( \lambda H(a)/k \right)^2}{1+\left( \lambda H(a)/k \right)^2} \right]
\label{muEvolution}
\end{equation}
and
\begin{equation}
\Sigma(a,k)=1+\Sigma_{0}\frac{\Omega_{\text{DE}}(a)}{\Omega_\Lambda} \left[ \frac{1+c_2 \left( \lambda H(a)/k \right)^2}{1+\left( \lambda H(a)/k \right)^2} \right].
\label{SigmaEvolution}
\end{equation}
Such a scale-dependent parameterization satisfies the property that at high-$k$ (small scales) $\Sigma-1 \rightarrow \Sigma_0 \Omega_{\text{DE}}(a)/\Omega_\Lambda$. However, for low-$k$ (large scales) we have $\Sigma-1 \rightarrow \Sigma_0 c_2 \Omega_{\text{DE}}(a)/\Omega_\Lambda$ and similarly for $\mu-1$ with $c_1$. Therefore, the parameters $c_1$ and $c_2$ characterize the behaviour of $\mu$ and $\Sigma$ at large scales, respectively. Here, the factor $H(a)/k$ is dimensionless since we are setting the speed of light as $c=1$. Also, we recover the redshift-dependence-only case when $\lambda=0$ or when $c_1=c_2=1$. 

\subsection{Binning methods \label{sec:binning-method}}
One advantage of assuming only redshift bins to model the MG parameters is that the number of extra parameters we need to fit is less than if we consider an additional scale dependence. However, previous efforts that only involved MG parameters in redshift bins found that there was not a significant improvement in the fit over $\Lambda$CDM and that there was good agreement with GR \cite{EMueller-2016}. Moreover, some works that used bins in both redshift and scale have found some mild to moderate tension in one of the small scale bins when fitting past surveys \cite{JohnsonEtAl2016,Daniel2010MG2,Joudaki2017-2,Joudaki2018}. Therefore, we opt to follow a similar procedure and use bins over the MG parameters in both redshift and scale. We choose to use four bins consisting of two bins in redshift and two bins in scale. We use the Integrated Software in Testing General Relativity (ISiTGR)~\cite{ISiTGR-CGQ,ISITGR}, which is an extension to CosmoMC~\cite{COSMOMC} and CAMB~\cite{CAMB} that includes MG phenomenological parameters. ISiTGR code already implements binning methods in the MG parameters relevant to the analysis we will perform. It has been shown in a comparative study ~\cite{Hamber2020} that among some of the current MG analysis packages, ISiTGR is the most reliable tool for testing the effects of a modified Newton’s constant in observational cosmology.
Using the notation given in the ISiTGR code, we consider the redshift bins to be described by $0<z\leq z_{\text{div}}$ and $z_{\text{div}}<z\leq z_{\text{TGR}}$ with $z_{\text{div}}=1$ and $z_{\text{TGR}}=2$, where $z_{\text{div}}$ is the redshift which splits the two bins and $z_{\text{TGR}}$ is the redshift above which GR is assumed. For the latter setting, the idea is that MG is present in the era of cosmic acceleration and that earlier in the universe gravity was governed by GR. This is consistent with the observation that cosmic acceleration only becomes dominant at a redshift of $\sim0.67$, so it is reasonable to assume GR for $z>z_{\text{TGR}}=2$. For the bins in scale we use $k\leq k_\text{c}$ and $k>k_\text{c}$ where $k_\text{c}=0.01$ Mpc$^{-1}$ is the scale which divides the bins. This value for $k_\text{c}$ roughly represents the scale at which the non-CMB probes start to play a role for $k>k_\text{c}$ and the matter-radiation equality horizon scale, which specifies the turnover scale in the matter power spectrum \cite{Prada_equality_horizon}. Then, if we represent each MG parameter by $X(z,k)$, the binned form of the MG parameter can be written as
\begin{equation}
\begin{split}
X(z,k) & = \frac{1+X_{z_1}(k)}{2}+\frac{X_{z_2}(k)-X_{z_1}(k)}{2}\tanh\left(\frac{z-z_{\text{div}}}{z_{\text{tw}}}\right) \\
& + \frac{1-X_{z_2}(k)}{2}\tanh\left(\frac{z-z_{\text{TGR}}}{z_{\text{tw}}}\right).
\end{split}
\label{eq:bins}
\end{equation}
This defines two redshift bins and then a "third bin" for GR after $z_{\text{TGR}}$, as shown in Table \ref{Table:Bins}.  
Here, expression (\ref{eq:bins}) gives a continuous transition of the MG parameters between redshift bins and $z_{\text{tw}}$ is a transition width parameter that sets how rapidly the transition from one bin to another occurs in time. A very small value of $z_{\text{tw}}$ might lead to numerical errors for bad MG parameter combinations and in such a case that combination in the parameter space would get rejected. In our case, we choose a moderate value for this transition parameter by using $z_{\text{tw}}=0.05$, which is hard-coded in ISiTGR but can still be easily modified in the code if needed. 

Now, we calculate the functions $X_{z_1}(k)$ and $X_{z_2}(k)$ using two different methods: The first is the traditional binning method, which consists of the use of a hyperbolic tangent function for scale bins as is done for the redshift bins. Here, the transition parameter for the scale bins is also hard-coded as $k_{\text{tw}}=k_{\text{c}}/10$. Hence, $X_{z_1}(k)$ and $X_{z_2}(k)$ are given by
\begin{equation}
X_{z_1}(k) = \frac{X_2+X_1}{2}+\frac{X_2-X_1}{2}\tanh\left(\frac{k-k_c}{k_{tw}}\right)
\end{equation}
and
\begin{equation}
X_{z_2}(k) = \frac{X_4+X_3}{2}+\frac{X_4-X_3}{2}\tanh\left(\frac{k-k_c}{k_{tw}}\right).
\end{equation}
We label this procedure as the traditional binning method. The second method to parameterize these functions is the hybrid method, in which the functions $X_{z_1}(k)$ and $X_{z_2}(k)$ follow a monotonic evolution inside each bin using exponential functions. Ultimately, the difference with respect to the traditional binning method is that hybrid binning leads to a smoother transition between the bins. In this case, the functions $X_{z_1}(k)$ and $X_{z_2}(k)$ are computed as 
\begin{equation}
X_{z_1}(k) = X_1 e^{-k/k_c} + X_2 (1-e^{-k/k_c})
\end{equation}
and
\begin{equation}
X_{z_2}(k) = X_3 e^{-k/k_c} + X_4 (1-e^{-k/k_c}).
\end{equation}
Thus, we study the evolution of the MG parameters by using both binning methods and show the ranges of the bins in Table \ref{Table:Bins}.
\begin{table}[t!]
\begin{center}
\scriptsize 
 \begin{tabular} {| c | c | c | c |}
\hline
  & \multicolumn{3}{c|}{Redshift bins} \\ \hline
Scale bins & $0\leq z< 1$ & $1\leq z< 2$ & $z\geq 2$ \\ \hline
$0\leq k < 0.01$ & $\mu_1$, $\Sigma_1$ & $\mu_3$, $\Sigma_3$ & GR is assumed\\ \hline
$0.01\leq k<\infty$ & $\mu_2$, $\Sigma_2$ & $\mu_4$, $\Sigma_4$ & GR is assumed \\
\hline
\end{tabular}
\end{center}
\caption{Redshift and scale bins used in this work, with the corresponding MG parameters. In the context of cosmic acceleration, deviations from GR are tested in the range $0 \leq z < 2$ (more discussion in the text). }
\label{Table:Bins}
\end{table}

\section{Datasets
\label{sec:datasets}}
In order to obtain meaningful constraints for MG parameters in each of the bins, it is important to combine different cosmological probes to which the MG parameters are sensitive. Hence we list below the current cosmological datasets used in our analysis that allow us to constrain the MG parameters along with the $\Lambda$CDM parameters.

\begin{itemize}
    \item[]{\textbf{Cosmic Microwave Background (CMB):} The CMB temperature and polarization measurements are a very powerful probe for the early universe and provide strong constraints on the primordial density perturbations as well as the MG parameters. CMB constitutes a major probe in constraining all the MG parameters we consider here since it provides constraints for quantities such as $C_{\ell}^{TT}$ or $C_{\ell}^{\phi\phi}$ which restrict the values of the MG parameters, especially in the low multipoles region of the spectra. Hence, we use the most recent CMB temperature and E-mode polarization measurements from the Planck mission. The Planck collaboration provides the temperature auto correlation measurements for the range $30 \leq\ell\leq 2058$, which we refer to as high-$\ell$, while for low-$\ell$ we use the temperature Commander likelihood. Additionally Planck provides the E-mode polarization auto correlation and the corresponding temperature-polarization cross correlation in the range $30 \leq\ell\leq 1996$ ~\cite{Planck2018}. We refer to the combined temperature and E-mode polarization spectra as TTTEEE. For the polarization in the low-$\ell$ region specified in the range $2 \leq\ell\leq 29$, we use the SimAll likelihood code for EE spectra, labeled as lowE. Additionally, the Planck release contains a likelihood associated with gravitational lensing measurements from the CMB, and we refer to it as CMBL ~\cite{Planck-2018-lensing}.} 
\end{itemize}

\begin{itemize}
    \item[]{\textbf{Weak lensing measurements:} We use the galaxy clustering and weak lensing measurements from the Dark Energy Survey (DES) Year 1 analysis ~\cite{DES2017}. We employ weak lensing and clustering data in our analysis from the $3\times 2$pt correlation functions which can produce tight constraints on $\Sigma$; this directly enters as a factor in the cosmic shear and galaxy-galaxy lensing measurements. Although less sensitive, $\mu$ can also be somewhat constrained by DES via galaxy clustering measurements. However, we restrict our use of the DES data to data points that correspond to linear scales, since our framework is only well defined at those scales. To remove the non-linear data points we consider a similar procedure as in other works ~\cite{MGCAMB3,DESMG2018,Planck2015MG}.}
\end{itemize}

\begin{itemize}
    \item[]{\textbf{Redshift space distortions:} A very informative cosmological probe for the growth rate of structure formation is given by Redshift Space Distortions (RSD). Basically, RSD is able to measure the combination $f(z)\sigma_8(z)$, where the growth factor of structure, $f(z)$, is affected by $\mu$, constituting a powerful probe for this MG parameter. We utilize the redshift space distortions measurements from the Baryon Oscillation Spectroscopic Survey (BOSS) Data Release 12 (DR12) which provides a combination of individual results into a set of consensus values and likelihoods in three effective redshifts: 0.38, 0.51 and 0.61 ~\cite{AlamEtAl2016}. We label this dataset as BAO/RSD.}
\end{itemize}

\begin{itemize}
    \item[]{\textbf{Background cosmological probes:} In order to break the degeneracies in the $\Lambda$CDM model at the background level we use Supernovae Type Ia (SNe) data and Baryon Acoustic Oscillations (BAO) measurements. We utilize the Pantheon sample data presented in ~\cite{Pan-STARRS-2017}. The Pantheon sample is based on a subset of 279 SNe discovered by Pan-STARRS1 at $0.03 < z < 0.68$, combined with useful distance estimations of 1048 SNe ranging from $0.01 < z < 2.3$ from other surveys. We also use BAO measurements from the consensus DR12 from BOSS at three effective redshifts, i.e. 0.38, 0.51 and 0.61 ~\cite{AlamEtAl2016}, in combination with BAO measurements from the Six Degree Field Galaxy Survey at $z_{\text{eff}}=0.106$ ~\cite{BeutlerEtAl2011} and the SDSS Data Release 7 Main Galaxy Sample at $z_{\text{eff}}=0.15$ ~\cite{RossEtAl2014}. Additionally, we include the quasar-Lyman-$\alpha$ cross-correlation and the Lyman-$\alpha$ forest auto-correlation measurements at $z_{\text{eff}}=2.35$ ~\cite{2019-BAO-lyalpha-quasar-Blomqvist}. We refer to the joint analysis of these measurements as the BAO dataset in this work.}
\end{itemize}
Finally, in order to simplify the notation used throughout this work, we refer to the combination TTTEEE+lowE as Planck and we assign the label SBR to the joint dataset SNe+BAO+ BAO/RSD.

\section{Results
\label{sec:results}}
In this section we report our results from performing a MG analysis based on functional form and binning methods. In order to do this we employ a Markov-Chain-Monte-Carlo (MCMC) procedure using ISiTGR for models based on the MG parameters, in addition to the six base $\Lambda$CDM cosmological parameters: the baryonic matter density $\Omega_bh^2$ and cold dark matter density $\Omega_c h^2$, the angular size of the sound horizon $\theta$, the reionization optical depth $\tau$, the spectral index $n_s$ and the amplitude of the primordial power spectrum $\ln(10^{10} A_s)$. We consider a general model with all the MG parameters as well as alternative models where some of the MG parameters are set to their GR values. This means that we alternatively do not consider either a gravitational slip, modifications in the growth of structure, or deviations in the bending of light predicted by GR. Thus, we constrain the MG parameters using ISiTGR as a MCMC sampler and then we obtain the associated statistics using GetDist \cite{Getdist}, which is a Python package for analysing and plotting MCMC samples. 

In the following, we present the constraints for the functional form given in Sec. \ref{sec:functional}. Afterwards, we focus on the MG parameter constraints using binning methods. Later on, we compare the constraints of the functional form and the binning methods in redshift and scale. Finally, we perform a model comparison with respect to the $\Lambda$CDM model.
\subsection{MG constraints using functional form for redshift and scale}

\begin{figure}[t]
\centering
\scriptsize 
\setlength{\tabcolsep}{2.5pt}
\begin{tabular}{c c}
  {\includegraphics[width=7.5cm]{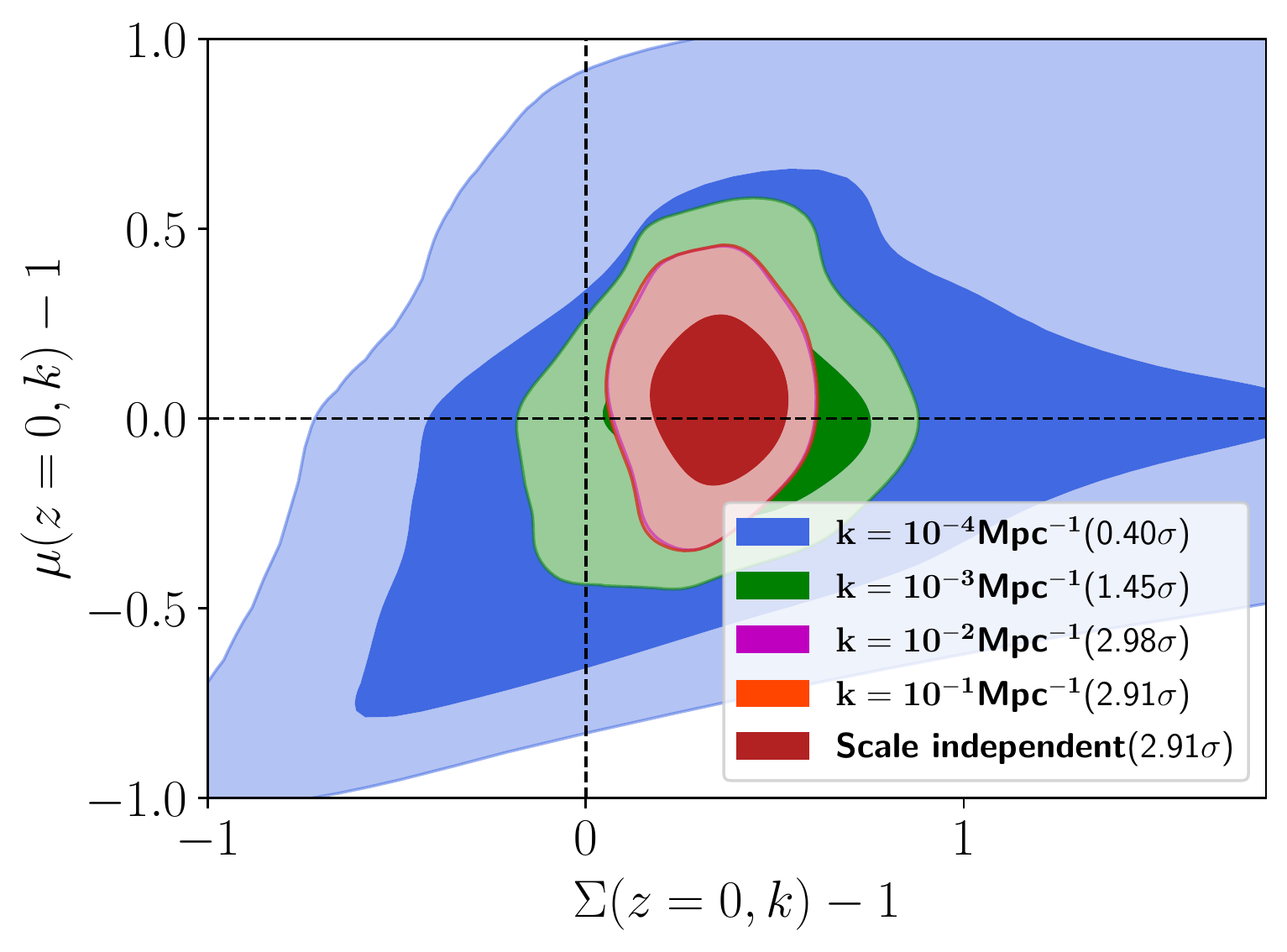}} & {\includegraphics[width=7.5cm]{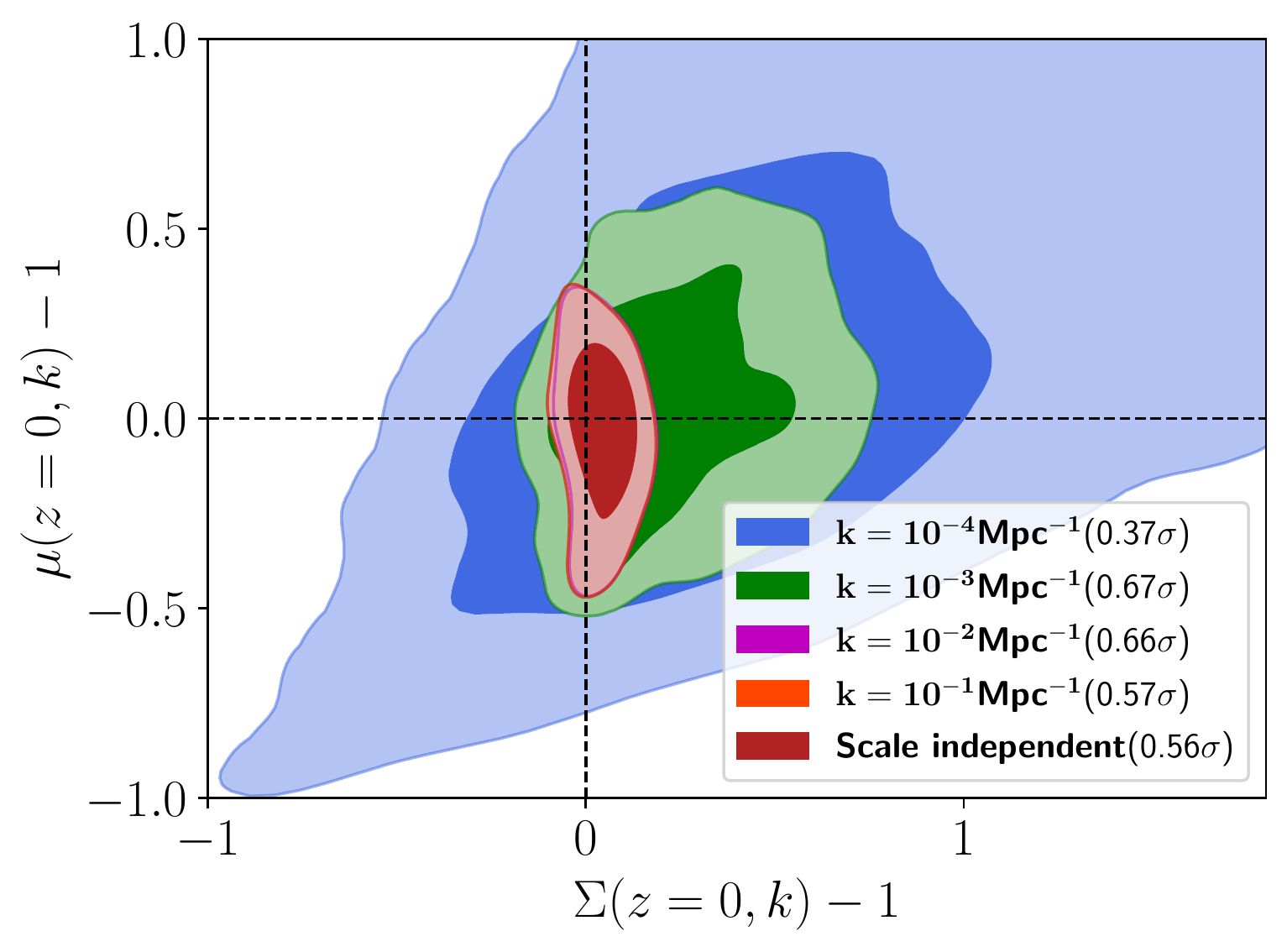}} \\
\end{tabular}
\caption{68\% and 95\% confidence contours for the functional form with scale-dependence shown in Sec. \ref{sec:functional} at a redshift $z = 0$. The left panel constraints use Planck+SBR while the right panel constraints use Planck+SBR+CMBL+DES. The red contours are obtained in the limit $k\rightarrow \infty$, which is the scale-independent limit of (\ref{muEvolution}) and (\ref{SigmaEvolution}). The dashed lines represent the GR value for each MG parameter, and the $n$-$\sigma$ value next to each scale regime denotes the tension with respect to GR. We can see that while there are notable tensions with GR for small scales using the Planck+SBR data, these tensions are greatly alleviated when we add lensing data. }
\label{Fig:Constraints_functional_twoMG}
\end{figure}
\begin{figure}[t]
\centering
\scriptsize 
\setlength{\tabcolsep}{2.5pt}
\begin{tabular}{c c}
  {\includegraphics[width=7cm]{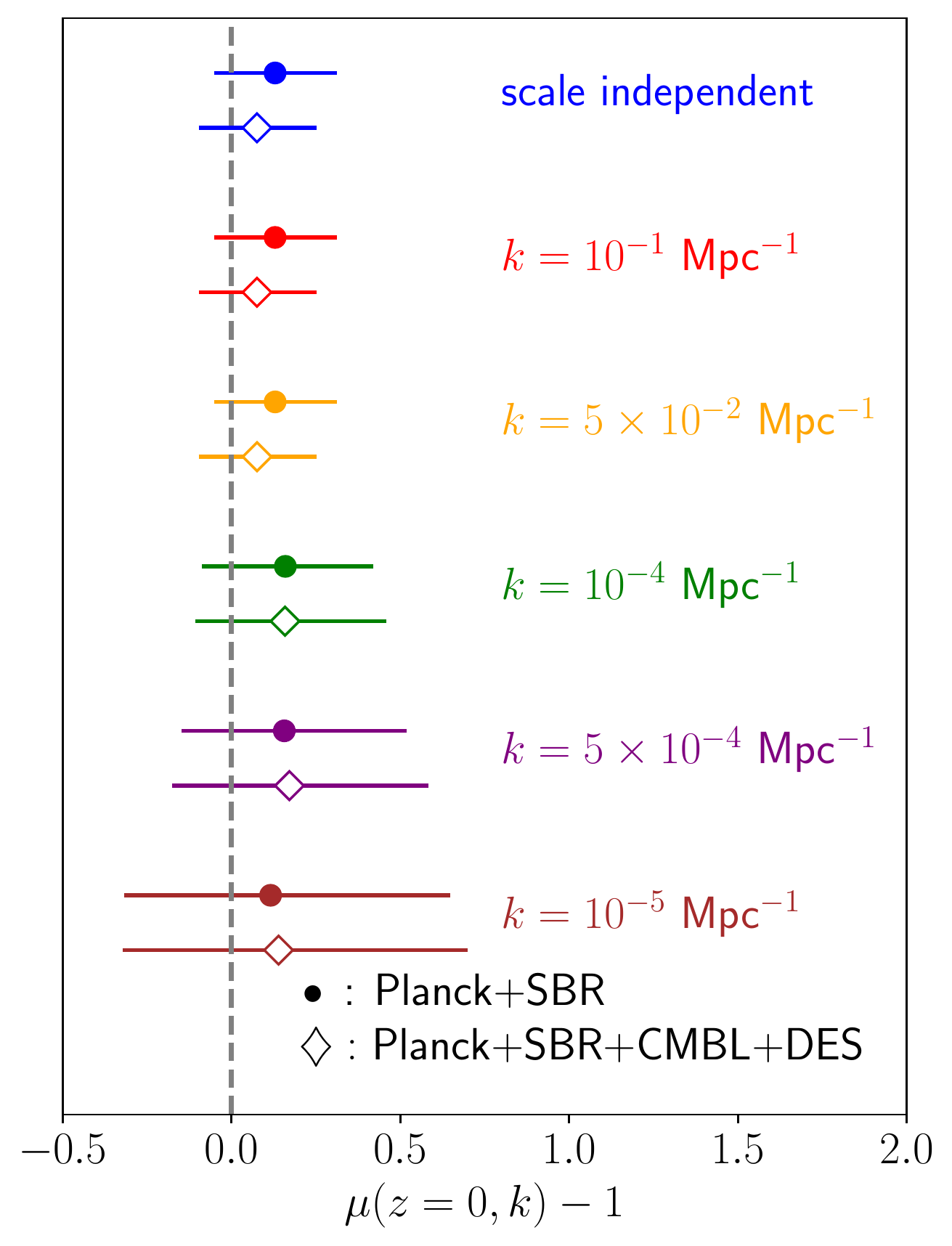}} & {\includegraphics[width=6.56cm]{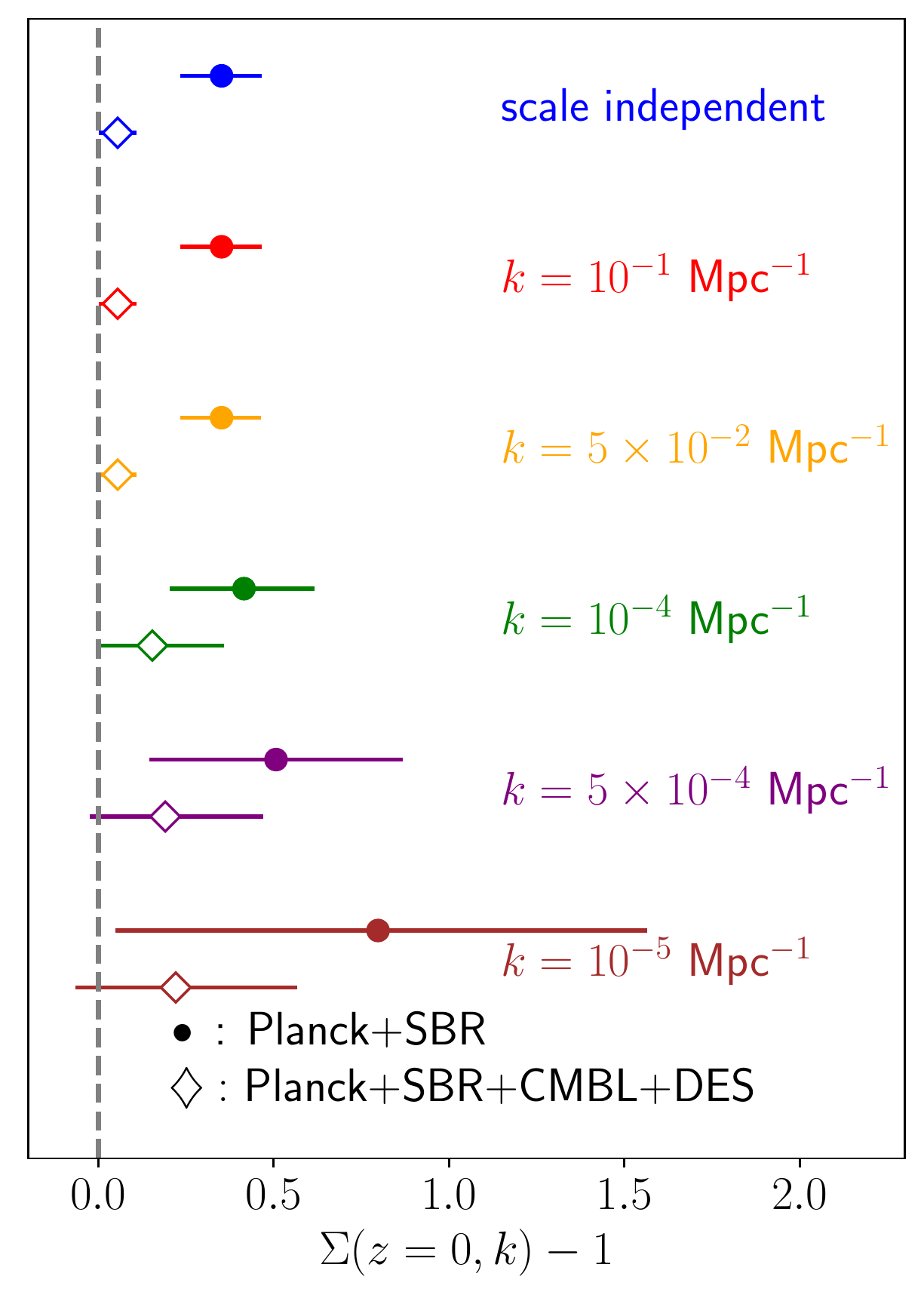}} \\
\end{tabular}
\caption{Mean values and 68\% limits for the functional form of the parameterizations $(\mu,\Sigma=1)$ and $(\mu =1, \Sigma)$, where $\Sigma=1$ in the left panel and $\mu=1$ in the right panel. The dashed lines represent the GR value for each MG parameter. Here, all the constraints in the left panel are consistent with GR, while in the right panel, $\Sigma(z=0,k)$ shows a 3.5-$\sigma$ tension with respect to GR when using Planck+SBR in the scale-independent case. This tension gets diluted to about 1-$\sigma$ as we consider larger scales since the error bars get larger. The tension also goes away when we add lensing data.}
\label{Fig:Constraints_functional_oneMG}
\end{figure}    

The functional form, with dependencies on redshift and scale, that we use for the MG parameters adds a total of five extra parameters to the six core cosmological parameters. When we set one MG parameter to its GR value we only add three extra parameters to $\Lambda$CDM (see equations \ref{muEvolution} and \ref{SigmaEvolution}). We are particularly interested here in analyzing the scale dependence of the MG parameter constraints and comparing it to other previous results with and without scale dependence. 

We show in Fig. \ref{Fig:Constraints_functional_twoMG} the constraints for the general case with two independent time- and scale-dependent MG parameters. We show the two-dimensional confidence contours for different values of $k$. Approximately, this scale dependence of the MG parameters starts to differ from the scale-independent scenario at larger scales such that $k<10^{-2}$Mpc$^{-1}$. When we use Planck+SBR we observe a tension at about 2.9-$\sigma$ for $k>10^{-2}\text{Mpc}^{-1}$. However, as we decrease the value of $k$, the volume of the 2-D contours increases. Therefore, below this scale limit we observe that the tension gets diluted as we consider larger scales.  A similar effect occurs when we consider Planck+SBR+CMBL+DES but in this case we do not observe any tension with GR for any scale value. Thus, we find that there is a tension at small scales when using Planck+SBR that is removed as we go below $k=10^{-2}\text{Mpc}^{-1}$, but we find no tension at any scale when lensing data is added. We point out that the broad contours at large scales are a consequence of the difficulty in constraining the parameters $c_1$, $c_2$ and $\lambda$, which are associated with the scale dependence of the MG parameters. Therefore, further scale-dependent parameterizations with fewer extra degrees of freedom might help to reduce the volume of the contours at large scales. 

Following up on this idea, it has been shown that higher tensions can be obtained by setting some MG parameters to their GR values \cite{CGQ_MG2020}, which reduces the number of extra degrees of freedom. Then, analogous constraints were obtained by setting $\mu_0=c_1=1$ or $\Sigma_0=c_2=1$ and they are shown in Fig. \ref{Fig:Constraints_functional_oneMG}. As we can observe, the constraints for $\mu(z=0,k)$ using Planck+SBR are similar to the constraints from Planck+SBR+CMBL+DES. This is expected since $\mu$ should not be very sensitive to the combination CMBL+DES, except for DES clustering measurements which do not provide strong constraints on $\mu$ and prefer a slightly negative value of this MG parameter \cite{DESMG2018}. On the other hand, we find that $\Sigma(z=0,k)$ is in some tension with GR when using Planck+SBR, but this tension gets alleviated as we observe larger scales. However, the combination Planck+SBR+CMBL+DES predicts a positive value of $\Sigma(z=0,k)$ but it is still statistically consistent with GR. Thus, only the combination Planck+SBR indicates a notable tension of 3.5-$\sigma$ for $\Sigma(z=0,k)$ when we observe small scales at $k\geq 10^{-2}\text{Mpc}^{-1}$.

\subsection{MG constraints using binning methods in redshift and scale}

\begin{table}[t!]
\begin{center}
\scriptsize 
 \begin{tabular} {| c | c | c | c |}
\hline
 Model & Parameterization & Binning parameters & Relationships \\ \hline
\textbf{P1} & ($\mu$, $\Sigma$) & $\mu_i$, $\Sigma_i$ & $\Sigma=\frac{\mu}{2}(\eta+1)$  \\ \hline
\textbf{P2} & ($\mu$, $\eta$=1) & $\mu_i$ & $\mu=\Sigma$ \\ \hline
\textbf{P3} & ($\mu$, $\Sigma$=1) & $\mu_i$ & $\mu=2/(\eta+1)$ \\ \hline
\textbf{P4} & ($\mu=1$, $\Sigma$) & $\Sigma_i$ & $\Sigma=(\eta+1)/2$ \\ 
\hline
\end{tabular}
\end{center}
\caption{Models used in the analysis with the corresponding parameterization and binning parameters. Only $\textbf{P1}$ is based on the full set of MG parameters, while the other models are particular cases where we fix some MG parameters. Finally, we list the relationship between MG parameters for each model.}
\label{Table:Models}
\end{table}
We perform a MG analysis based on binning methods in both redshift and scale using the current data. We aim to analyze the constraints to see how the tension with respect to GR is distributed within the bins and compare our results to functional forms. We consider four MG models as listed in Table \ref{Table:Models}: We consider a general model with two independent MG parameters as well as other models with one fixed MG parameter and one freely varying parameter, and employ bins to parameterize each of the independent parameters in all models. We consider four bins per each MG parameter, so in addition to the six core cosmological parameters we have a total of eight extra parameters for model \textbf{P1}, and four extra parameters for the rest of the MG models (i.e. \textbf{P2}-\textbf{P4}).

 We show in Table \ref{Table:Constraints_2MG} our results for the mean values and the associated uncertainties for the model \textbf{P1}, where we have varied the full set of MG parameters in addition to $\Lambda$CDM parameters. As we can observe, overall the $\mu_i$ parameters seem to be in agreement with GR when we consider the traditional binning. However, the hybrid binning indicates some tension in the marginalized 1-D constraints for the parameters $\mu_2$ and $\mu_4$ when we consider the dataset Planck+SBR+CMBL+DES. We observe that this 2-$\sigma$ tension is not shown by the traditional approach which indicated that all $\mu_i$ parameters are in agreement with their GR value of 1 with no tension above the 95\% level. Moreover, this 2-$\sigma$ tension is not present when we consider the 2-D confidence contours for the $\mu_i$ parameters.
 
\begin{table}[t!]
\setlength{\tabcolsep}{2pt}
\scriptsize 
\begin{center}
\begin{tabular}{| p{2cm} c|c|c|c|c|c|c|c|c|}
\multicolumn{1}{c}{} \\ \hline
\multirow{2}{*}{Datasets} & & \multicolumn{4}{c}{ Traditional binning} & \multicolumn{4}{|c|}{ Hybrid binning}\\ \cline{3-10}
& & $i=1$ & $i=2$ & $i=3$ & $i=4$ & $i=1$ & $i=2$ & $i=3$ & $i=4$ \\ \hline \hline
\multirow{2}{*}{Planck+BAO} & \multicolumn{1}{|c|}{$\mu_i$} & $1.08^{+0.94}_{-1.1}       $ & $1.6^{+1.2}_{-1.0}         $ & $1.02^{+0.80}_{-0.85}      $ & $0.53^{+0.83}_{-0.90}      $ & $1.0^{+1.0}_{-1.2}         $ & $1.6^{+1.2}_{-1.1}         $ & $1.06^{+0.82}_{-0.99}      $ & $0.53^{+0.90}_{-0.90}      $ \\ \cline{2-10}
/RSD & \multicolumn{1}{|c|}{$\Sigma_i$} & $1.04^{+0.10}_{-0.10}      $ & $\mathbf{1.19^{+0.13}_{-0.14}}      $ & $1.014^{+0.056}_{-0.057}   $ & $\mathbf{1.087^{+0.077}_{-0.084}}   $ & $0.98^{+0.20}_{-0.19}      $ & $\mathbf{1.19^{+0.16}_{-0.17}}      $ & $0.97^{+0.11}_{-0.10}      $ & $\mathbf{1.12^{+0.10}_{-0.11}}      $ \\ \hline \hline
\multirow{2}{*}{Planck+SBR} & \multicolumn{1}{|c|}{$\mu_i$} & $1.08^{+0.95}_{-1.1}       $ & $1.6^{+1.2}_{-1.1}         $ & $1.07^{+0.77}_{-0.84}      $ & $0.58^{+0.90}_{-0.92}      $ & $0.98^{+1.0}_{-1.3}        $ & $1.6^{+1.2}_{-1.1}         $ & $1.06^{+0.80}_{-0.97}      $ & $0.56^{+0.87}_{-0.89}      $ \\ \cline{2-10}
 & \multicolumn{1}{|c|}{$\Sigma_i$} & $1.04^{+0.10}_{-0.10}      $ & $\mathbf{1.19^{+0.12}_{-0.15}}      $ & $1.014^{+0.057}_{-0.056}   $ & $\mathbf{1.084^{+0.075}_{-0.084}}   $ & $0.98^{+0.19}_{-0.18}      $ & $\mathbf{1.20^{+0.17}_{-0.18}}      $ & $0.96^{+0.11}_{-0.10}      $ & $\mathbf{1.12^{+0.10}_{-0.11}}      $ \\ \hline \hline
\multirow{2}{*}{Planck+SBR} & \multicolumn{1}{|c|}{$\mu_i$} & $1.07^{+0.99}_{-1.1}       $ & $1.9^{+1.2}_{-1.1}         $ & $1.01^{+0.80}_{-0.85}      $ & $0.33^{+0.88}_{-0.83}      $ & $0.98^{+1.1}_{-1.2}        $ & $1.8^{+1.2}_{-1.1}         $ & $1.03^{+0.88}_{-1.0}       $ & $0.40^{+0.91}_{-0.91}      $ \\ \cline{2-10}
+CMBL & \multicolumn{1}{|c|}{$\Sigma_i$} & $1.04^{+0.10}_{-0.11}      $ & $\mathbf{1.16^{+0.13}_{-0.14} }     $ & $1.017^{+0.053}_{-0.059}   $ & $1.068^{+0.074}_{-0.086}   $ & $1.03^{+0.19}_{-0.19}      $ & $1.14^{+0.15}_{-0.15}      $ & $0.99^{+0.11}_{-0.11}      $ & $1.08^{+0.10}_{-0.11}      $ \\ \hline \hline
 \multirow{2}{*}{Planck+SBR} & \multicolumn{1}{|c|}{$\mu_i$} & $1.04^{+0.98}_{-1.2}       $ & $1.9^{+1.2}_{-1.1}         $ & $1.08^{+0.75}_{-0.82}      $ & $0.23^{+0.86}_{-0.87}      $ & $0.9^{+1.1}_{-1.3}         $ & $2.0^{+1.1}_{-1.1}         $ & $1.27^{+0.64}_{-0.77}      $ & $\mathbf{0.14^{+0.85}_{-0.79}}      $ \\ \cline{2-10}
+DES & \multicolumn{1}{|c|}{$\Sigma_i$} & $1.04^{+0.10}_{-0.10}      $ & $1.08^{+0.12}_{-0.12}      $ & $1.014^{+0.058}_{-0.056}   $ & $\mathbf{1.086^{+0.072}_{-0.080}}   $ & $1.06^{+0.16}_{-0.16}      $ & $1.08^{+0.13}_{-0.12}      $ & $0.96^{+0.11}_{-0.10}      $ & $\mathbf{1.13^{+0.11}_{-0.11}  }    $ \\ \hline \hline
\multirow{2}{*}{Planck+SBR} & \multicolumn{1}{|c|}{$\mu_i$} & $1.0^{+1.0}_{-1.2}         $ & $1.87^{+1.1}_{-0.99}       $ & $1.13^{+0.73}_{-0.80}      $ & $0.26^{+0.82}_{-0.76}      $ & $0.9^{+1.2}_{-1.5}         $ & $\mathbf{2.04^{+1.1}_{-0.99}   }    $ & $1.25^{+0.67}_{-0.82}      $ & $\mathbf{0.14^{+0.77}_{-0.80}      }$ \\ \cline{2-10}
+CMBL+DES & \multicolumn{1}{|c|}{$\Sigma_i$} & $1.04^{+0.10}_{-0.11}      $ & $1.08^{+0.12}_{-0.11}      $ & $1.013^{+0.057}_{-0.059}   $ & $\mathbf{1.088^{+0.062}_{-0.070} }  $ & $1.08^{+0.17}_{-0.18}      $ & $1.08^{+0.12}_{-0.12}      $ & $0.974^{+0.098}_{-0.091}   $ & $\mathbf{1.116^{+0.087}_{-0.095} }  $ \\ \hline
\end{tabular}
\end{center}
\caption{Marginalized means and 95\% confidence limits using several datasets for model \textbf{P1}. We model the MG parameters using both the traditional and hybrid binning methods. The boldface values indicate the constraints that are in a higher than 2-$\sigma$ tension with respect to GR. Again, the subscripts on the MG parameters correspond to their bins in redshift and scale, as in Table \ref{Table:Bins}. We observe that overall the $\mu_i$ parameters agree well with GR (with the exception of $\mu_2$ and $\mu_4$ with hybrid binning using Planck+SBR+CMBL+DES). However, there is a variety of tensions when we examine $\Sigma_2$ and $\Sigma_4$ across different binning and dataset regimes; although $\Sigma_2$ can be alleviated in some cases with the addition of lensing data, the tension in $\Sigma_4$ remains in almost all cases.}
\label{Table:Constraints_2MG}
\end{table}
\begin{figure}[t]
\centering
\scriptsize 
\setlength{\tabcolsep}{2.5pt}
\begin{tabular}{c c}
  {\includegraphics[width=7cm]{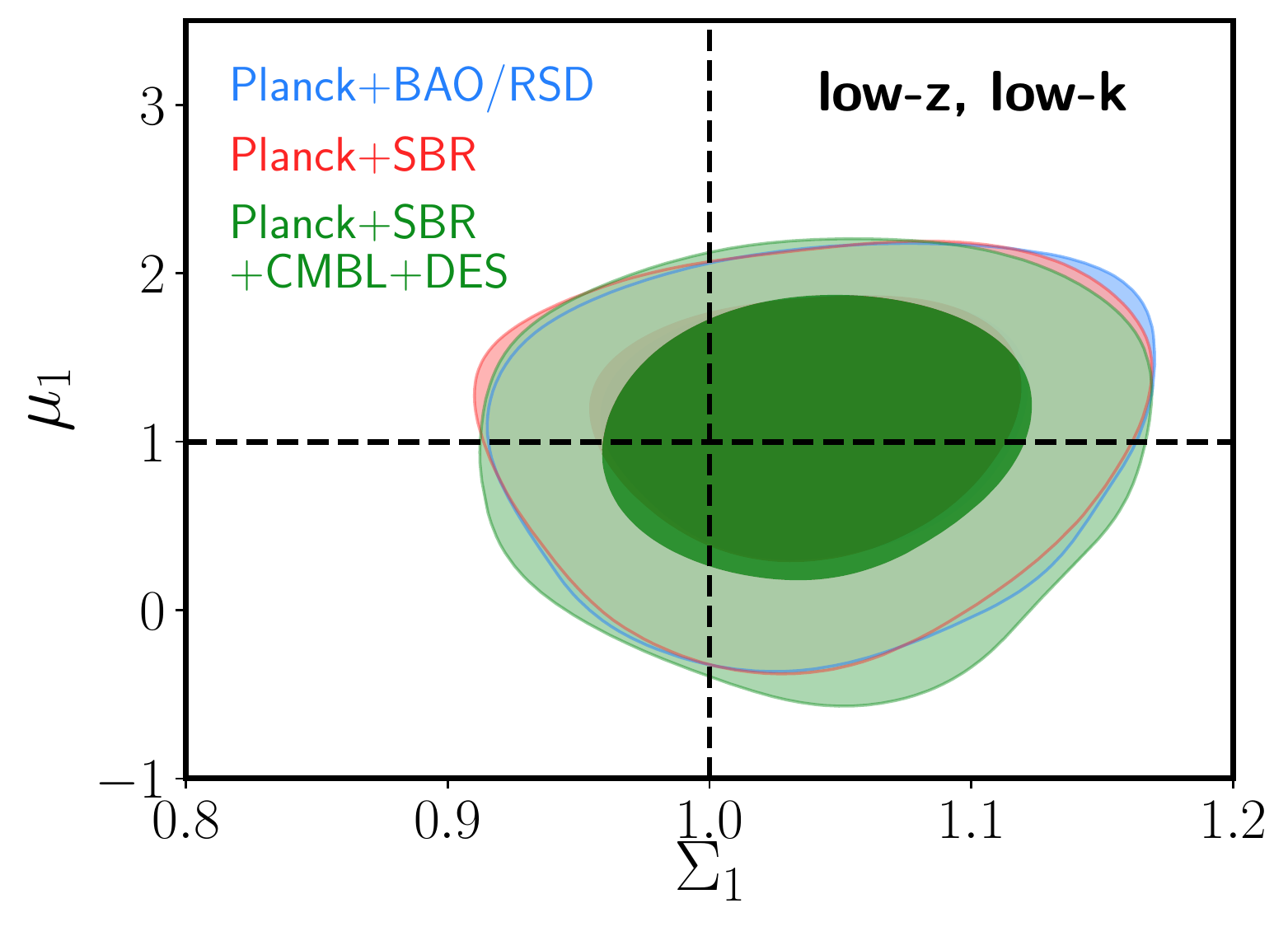}} & {\includegraphics[width=7cm]{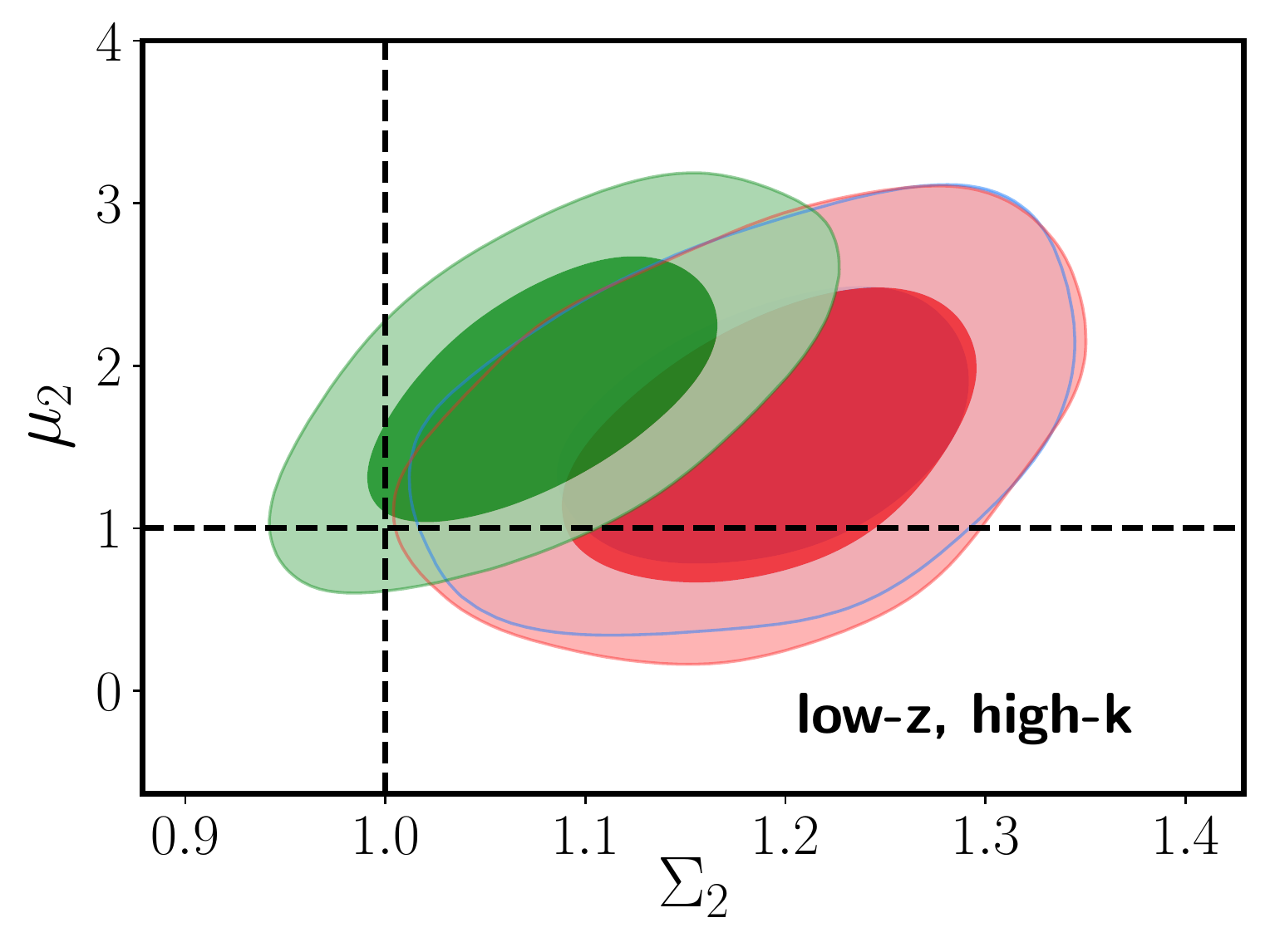}} \\
  {\includegraphics[width=7cm]{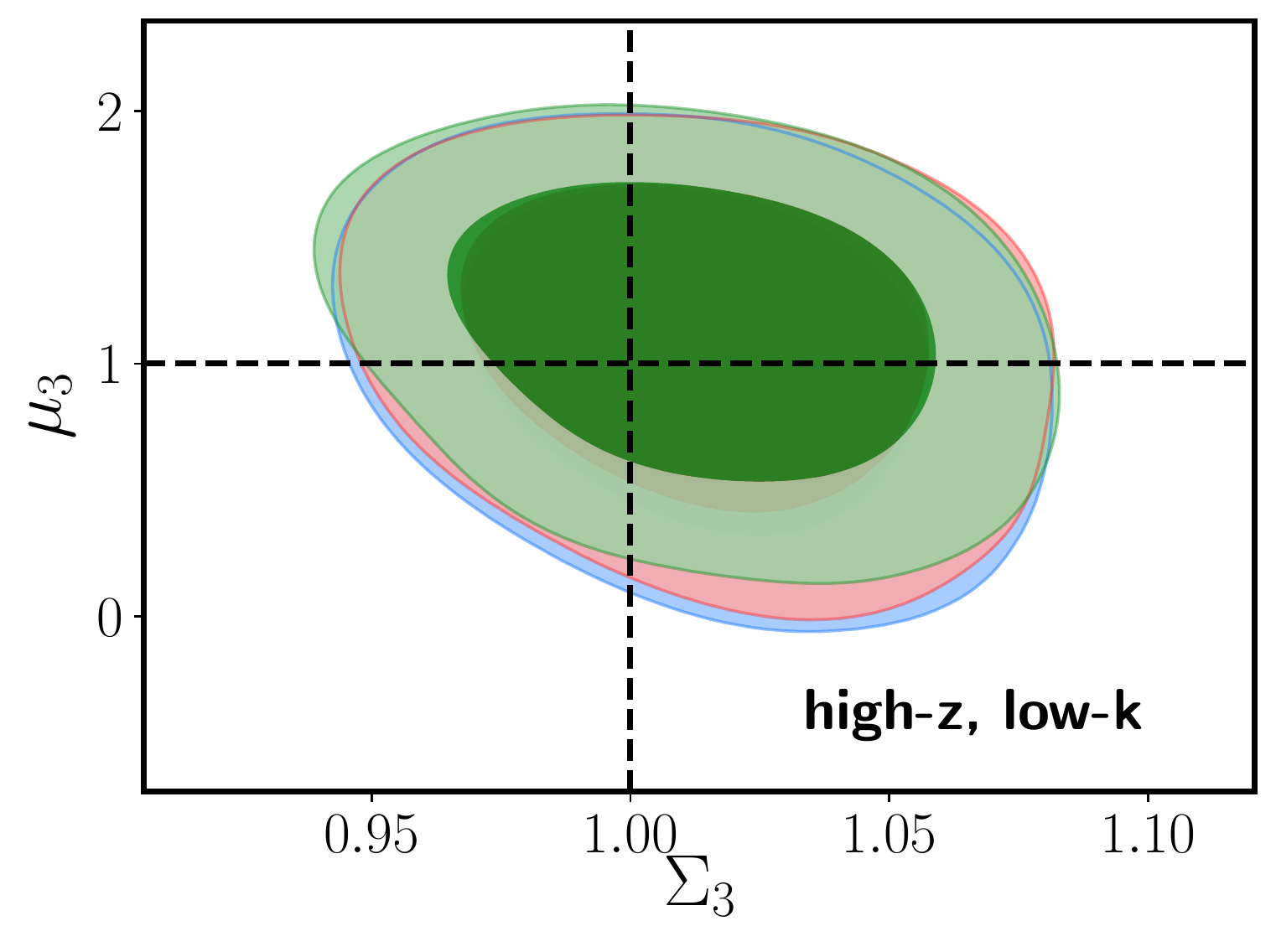}} & {\includegraphics[width=7cm]{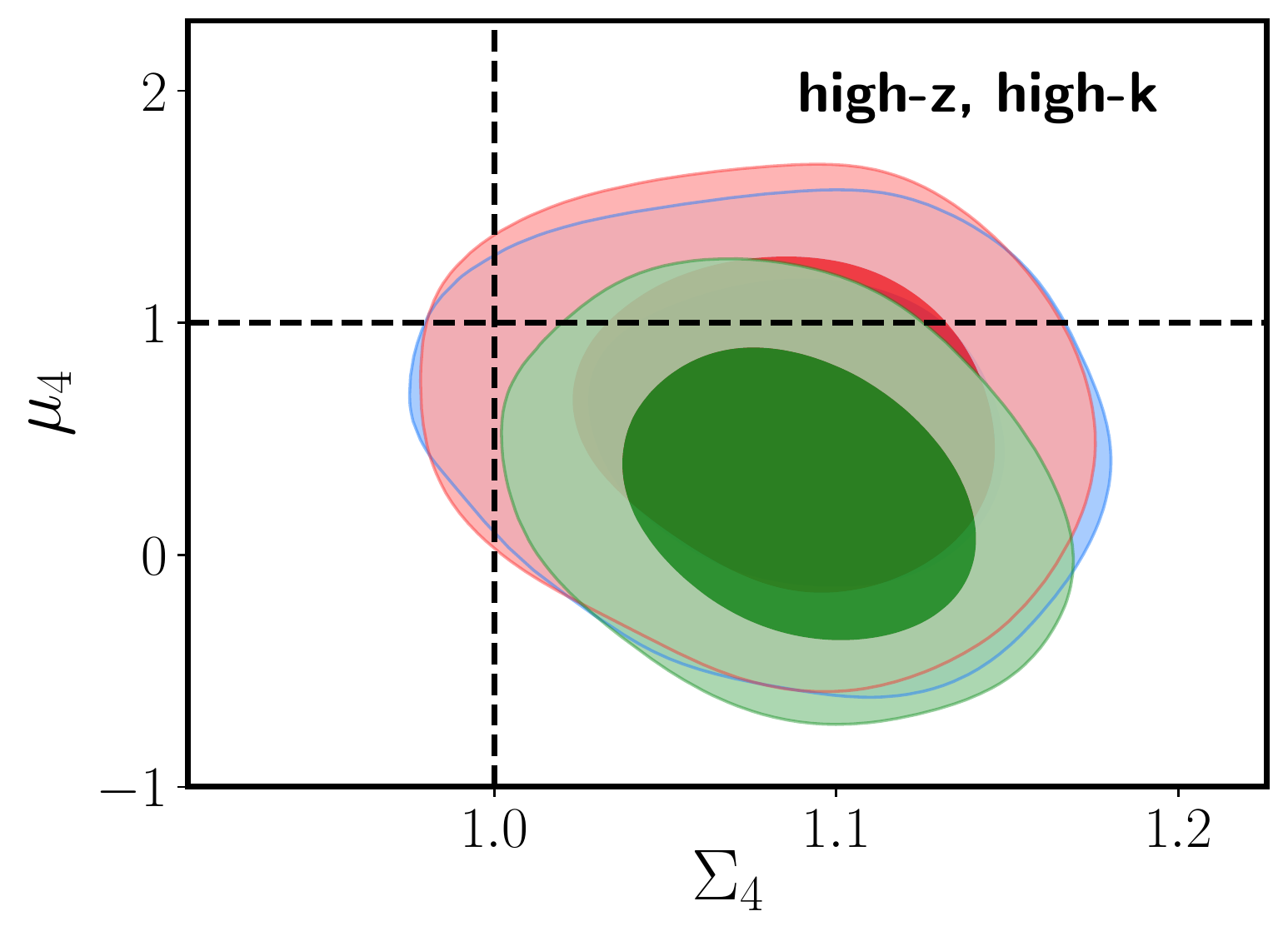}} \\
\end{tabular}
\caption{68\% and 95\% confidence contours for the constraints obtained for model \textbf{P1} using the traditional binning method. We form four different combinations of MG parameters that correspond to the four different bins in redshift and scale (see Table \ref{Table:Bins}). The dashed lines show the GR values. We observe again that there is noticeable tension in $\Sigma_2$ and $\Sigma_4$, and while adding lensing data reduces $\Sigma_2$'s tension, we still observe a 2.6-$\sigma$ tension in $\Sigma_4$.}
\label{Fig:Constraints_2MG}
\end{figure}

\begin{table}[t!]
\setlength{\tabcolsep}{4pt}
\scriptsize 
\begin{center}
\begin{tabular}{| p{3cm} | c | c | c | c | c | c | c | c |}
 \multicolumn{1}{c}{} & \multicolumn{1}{c}{} &  \multicolumn{1}{c}{} &  \multicolumn{1}{c}{} &  \multicolumn{1}{c}{} &  \multicolumn{1}{c}{} &  \multicolumn{1}{c}{} \\ \hline
Datasets & \multicolumn{2}{c|}{\textbf{P2:} trad} & \textbf{P2:} hybrid & \textbf{P3:} trad & \textbf{P3:} hybrid & \multicolumn{2}{c|}{\textbf{P4:} trad} & \textbf{P4:} hybrid \\ \hline \hline
\multirow{4}{*}{Planck+BAO/RSD} & $\mu_1$ & $1.042^{+0.088}_{-0.080}   $ & $1.02^{+0.17}_{-0.16}      $ & $1.00^{+0.94}_{-1.1}       $ & $0.98^{+1.0}_{-1.2}        $ & $\Sigma_1$ & $1.038^{+0.088}_{-0.083}   $ & $0.99^{+0.19}_{-0.16}      $ \\ \cline{2-9}
& $\mu_2$ & $\mathbf{1.13^{+0.12}_{-0.12}}      $ & $1.12^{+0.14}_{-0.15}      $ & $1.07^{+1.1}_{-0.96}       $ & $1.04^{+1.0}_{-0.98}       $ & $\Sigma_2$ & $\mathbf{1.16^{+0.12}_{-0.13}}      $ & $1.16^{+0.15}_{-0.17}      $ \\ \cline{2-9}
& $\mu_3$ & $1.015^{+0.056}_{-0.053}   $ & $0.985^{+0.11}_{-0.099}    $ & $1.21^{+0.66}_{-0.71}      $ & $1.19^{+0.75}_{-0.84}      $ & $\Sigma_3$ & $1.013^{+0.054}_{-0.054}   $ & $0.97^{+0.10}_{-0.10}      $ \\ \cline{2-9}
& $\mu_4$ & $1.063^{+0.080}_{-0.084}   $ & $1.08^{+0.10}_{-0.11}      $ & $0.99^{+0.82}_{-0.84}      $ & $1.04^{+0.80}_{-0.85}      $ & $\Sigma_4$ & $1.078^{+0.075}_{-0.082}   $ & $1.11^{+0.11}_{-0.11}      $ \\ \hline \hline
 \multirow{4}{*}{Planck+SBR} &  $\mu_1$ & $1.042^{+0.090}_{-0.082}   $ & $1.01^{+0.18}_{-0.15}      $ & $1.01^{+0.92}_{-1.0}       $ & $0.98^{+1.0}_{-1.2}        $ & $\Sigma_1$ & $1.037^{+0.087}_{-0.085}   $ & $0.99^{+0.18}_{-0.15}      $ \\ \cline{2-9}
&  $\mu_2$ & $\mathbf{1.14^{+0.12}_{-0.13} }     $ & $1.13^{+0.14}_{-0.15}      $ & $1.1^{+1.0}_{-1.0}         $ & $0.97^{+1.0}_{-0.96}       $ & $\Sigma_2$ & $\mathbf{1.16^{+0.11}_{-0.13}}      $ & $\mathbf{1.17^{+0.14}_{-0.16}    }  $ \\ \cline{2-9}
 &  $\mu_3$ & $1.015^{+0.055}_{-0.053}   $ & $0.99^{+0.11}_{-0.10}      $ & $1.23^{+0.66}_{-0.70}      $ & $1.16^{+0.76}_{-0.87}      $ & $\Sigma_3$ & $1.012^{+0.055}_{-0.054}   $ & $0.962^{+0.10}_{-0.098}    $ \\ \cline{2-9}
& $\mu_4$ & $1.066^{+0.074}_{-0.082}   $ & $1.09^{+0.10}_{-0.11}      $ & $1.00^{+0.83}_{-0.84}      $ & $1.10^{+0.84}_{-0.83}      $ & $\Sigma_4$ & $\mathbf{1.083^{+0.074}_{-0.082}}   $ & $\mathbf{1.12^{+0.10}_{-0.11}  }    $ \\ \hline \hline
 \multirow{4}{*}{Planck+SBR} & $\mu_1$ & $1.042^{+0.086}_{-0.085}   $ & $1.04^{+0.17}_{-0.16}      $ & $1.00^{+0.93}_{-1.1}       $ & $0.996^{+0.99}_{-1.2}      $ & $\Sigma_1$ & $1.042^{+0.088}_{-0.085}   $ & $1.04^{+0.18}_{-0.16}      $ \\ \cline{2-9}
&  $\mu_2$ & $1.09^{+0.10}_{-0.11}      $ & $1.08^{+0.12}_{-0.13}      $ & $0.90^{+0.92}_{-0.84}      $ & $0.88^{+0.87}_{-0.80}      $ & $\Sigma_2$ & $1.099^{+0.095}_{-0.10}    $ & $1.09^{+0.12}_{-0.12}      $ \\ \cline{2-9}
 \multirow{1}{*}{+CMBL} & $\mu_3$ & $1.015^{+0.055}_{-0.056}   $ & $1.008^{+0.097}_{-0.099}   $ & $1.24^{+0.65}_{-0.70}      $ & $1.19^{+0.76}_{-0.85}      $ & $\Sigma_3$ & $1.015^{+0.056}_{-0.057}   $ & $0.997^{+0.10}_{-0.10}     $ \\ \cline{2-9}
&  $\mu_4$ & $1.039^{+0.077}_{-0.075}   $ & $1.05^{+0.10}_{-0.10}      $ & $1.16^{+0.69}_{-0.72}      $ & $1.17^{+0.69}_{-0.71}      $ & $\Sigma_4$ & $1.046^{+0.078}_{-0.084}   $ & $1.06^{+0.10}_{-0.11}      $ \\ \hline \hline
 \multirow{4}{*}{Planck+SBR}&  $\mu_1$ & $1.044^{+0.086}_{-0.081}   $ & $1.12^{+0.14}_{-0.14}      $ & $0.99^{+0.95}_{-1.1}       $ & $0.9^{+1.1}_{-1.2}         $ & $\Sigma_1$ & $1.041^{+0.093}_{-0.085}   $ & $1.10^{+0.14}_{-0.15}      $ \\ \cline{2-9}
&  $\mu_2$ & $1.002^{+0.083}_{-0.077}   $ & $0.985^{+0.090}_{-0.085}   $ & $1.50^{+1.0}_{-0.96}       $ & $1.57^{+0.92}_{-0.90}      $ & $\Sigma_2$ & $1.016^{+0.087}_{-0.087}   $ & $1.007^{+0.090}_{-0.090}   $ \\ \cline{2-9}
 \multirow{1}{*}{+DES}&  $\mu_3$ & $1.015^{+0.053}_{-0.054}   $ & $0.98^{+0.11}_{-0.10}      $ & $1.27^{+0.63}_{-0.69}      $ & $1.42^{+0.58}_{-0.68}      $ & $\Sigma_3$ & $1.013^{+0.055}_{-0.054}   $ & $0.962^{+0.11}_{-0.096}    $ \\ \cline{2-9}
&  $\mu_4$ & $1.061^{+0.074}_{-0.082}   $ & $1.084^{+0.096}_{-0.10}    $ & $0.59^{+0.73}_{-0.70}      $ & $0.53^{+0.66}_{-0.65}      $ & $\Sigma_4$ & $1.076^{+0.076}_{-0.083}   $ & $\mathbf{1.112^{+0.092}_{-0.10}}    $ \\ \hline \hline
\multirow{4}{*}{Planck+SBR}& $\mu_1$ & $1.047^{+0.083}_{-0.083}   $ & $1.12^{+0.14}_{-0.15}      $ & $0.9^{+1.0}_{-1.0}         $ & $0.9^{+1.0}_{-1.2}         $ & $\Sigma_1$ & $1.047^{+0.089}_{-0.085}   $ & $1.11^{+0.14}_{-0.15}      $ \\ \cline{2-9}
&  $\mu_2$ & $1.009^{+0.081}_{-0.070}   $ & $0.989^{+0.082}_{-0.080}   $ & $1.25^{+0.85}_{-0.86}      $ & $1.30^{+0.83}_{-0.85}      $ & $\Sigma_2$ & $1.017^{+0.080}_{-0.072}   $ & $1.001^{+0.086}_{-0.082}   $ \\ \cline{2-9}
 \multirow{1}{*}{+CMBL+DES}&  $\mu_3$ & $1.019^{+0.052}_{-0.053}   $ & $0.993^{+0.10}_{-0.096}    $ & $1.33^{+0.60}_{-0.66}      $ & $1.42^{+0.58}_{-0.70}      $ & $\Sigma_3$ & $1.017^{+0.055}_{-0.055}   $ & $0.984^{+0.10}_{-0.097}    $ \\ \cline{2-9}
&  $\mu_4$ & $1.057^{+0.067}_{-0.077}   $ & $1.073^{+0.090}_{-0.097}   $ & $0.80^{+0.65}_{-0.65}      $ & $0.75^{+0.62}_{-0.60}      $ & $\Sigma_4$ & $1.071^{+0.064}_{-0.073}   $ & $1.091^{+0.089}_{-0.095}   $ \\ \hline
 \end{tabular}
\end{center}
\caption{Expectation values and 95\% limits for the MG parameters using the traditional and hybrid binning methods. We only show here the constraints for the models where we have set some MG parameters equal to their GR value (i.e. models \textbf{P2}, \textbf{P3}, and \textbf{P4}, see Table \ref{Table:Models}). MG parameter values that are in a tension greater than 2-$\sigma$ with respect to GR are shown in bold. Notable tensions include $\mu_2$ in model \textbf{P2} ($\eta = 1)$ with the traditional binning method and no lensing data, and a variety of $\Sigma_2$ and $\Sigma_4$ tensions in model \textbf{P4} ($\mu = 1$) across the different binning methods, mostly for datasets with no lensing. We also point out that there are no tensions $>$ 2-$\sigma$ for the model \textbf{P3}, where we set $\Sigma = 1$. See more discussion in the main text.}
\label{Table:Constraints_1MG}
\end{table}

\begin{figure}[t]
\centering
\setlength{\tabcolsep}{-2pt}
\begin{tabular}{c c c}
  {\includegraphics[width=5.3cm]{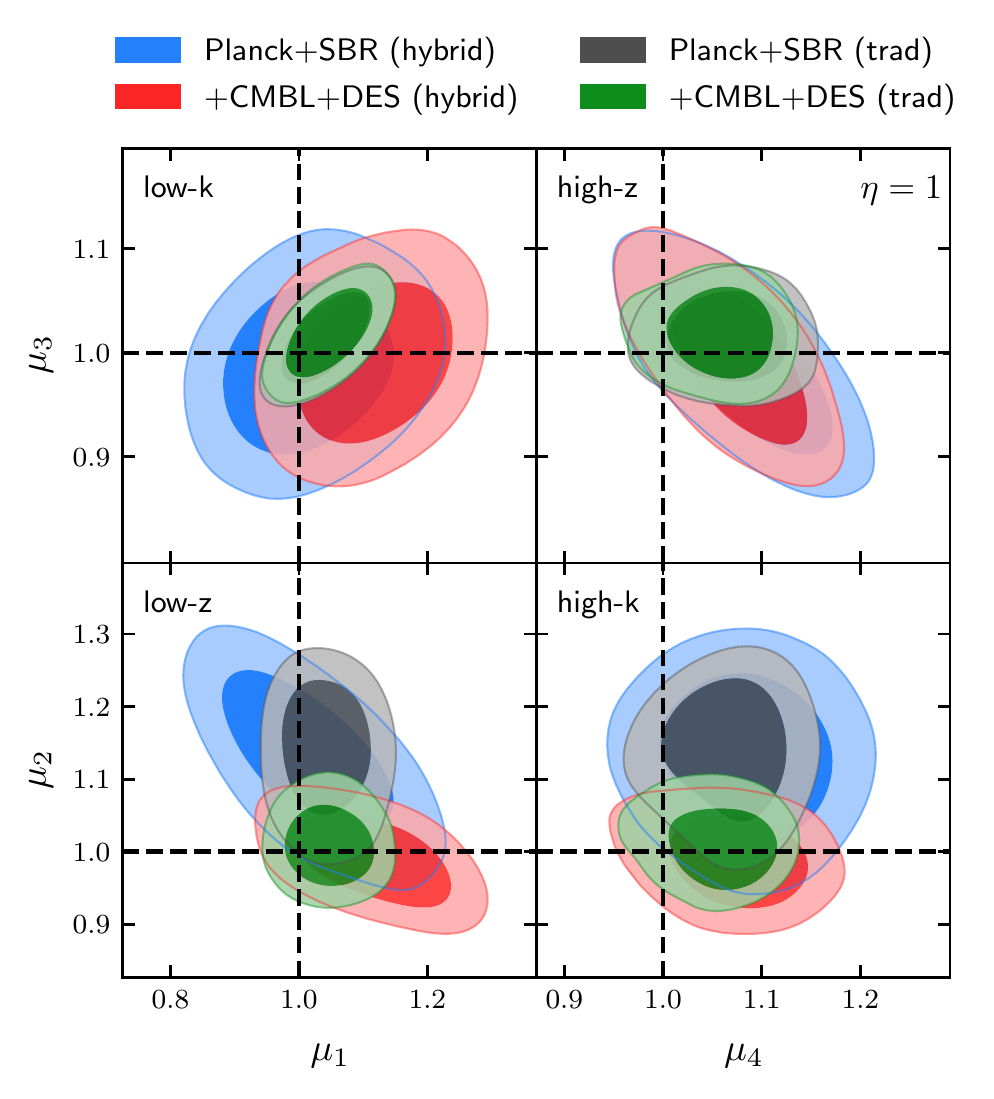}} &
  {\includegraphics[trim={0.1cm 0.1cm 0.1cm 0.1cm},clip, width=5.3cm]{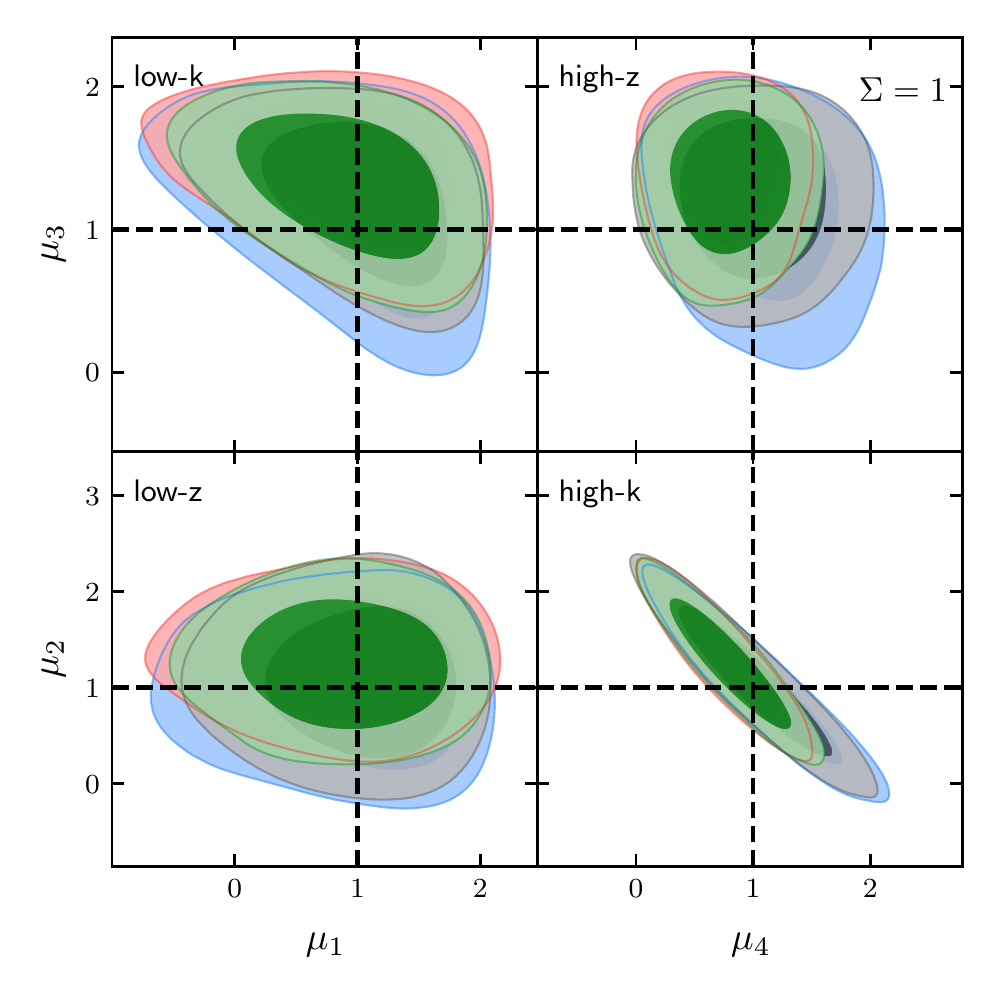}} &
  {\includegraphics[trim={0.1cm 0.1cm 0.1cm 0.1cm},clip, width=5.3cm]{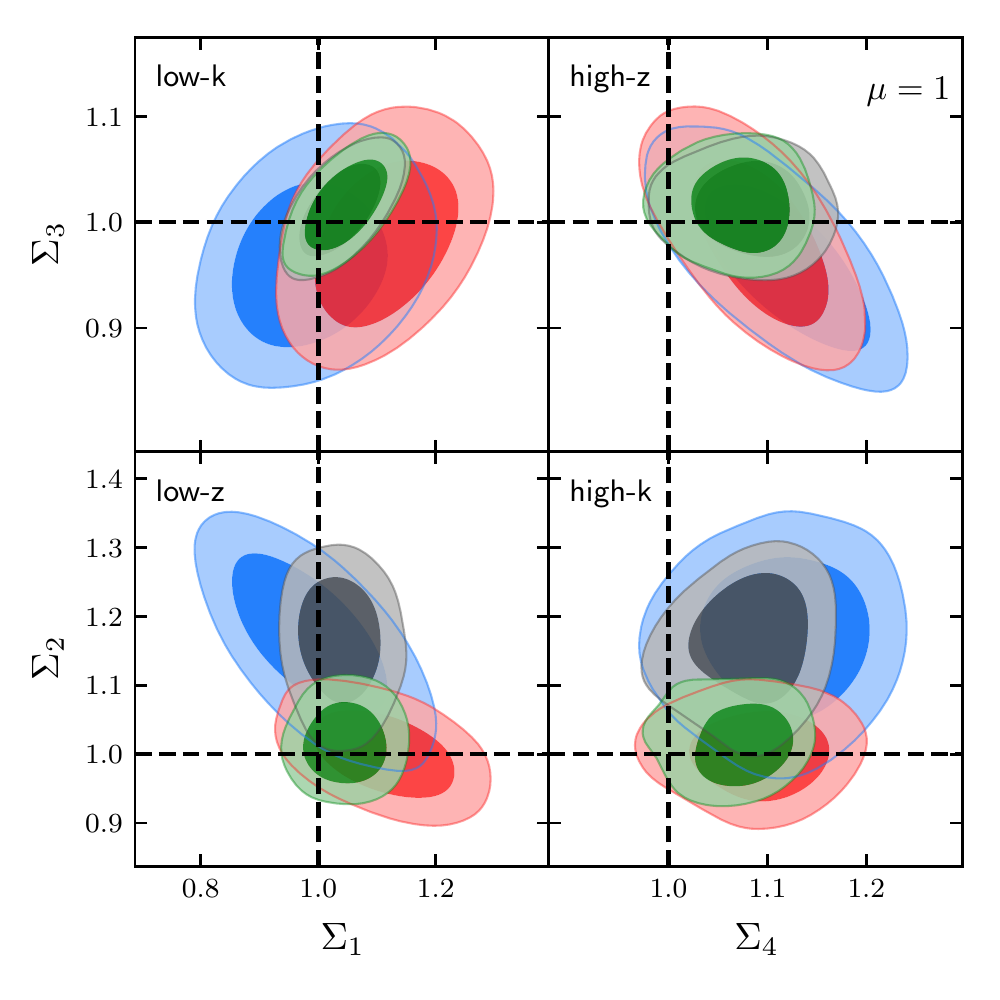}} \\
\end{tabular}
\caption{68\% and 95\% confidence contours for the \textbf{P2} (left panel), \textbf{P3} (middle panel) and \textbf{P4} models (right panel) where we have set one MG parameter to its GR value. We show the constraints in a rectangular plot way, which better illustrates the constraints on each redshift or scale bin. For example, a two-dimensional plot for $\mu_1$ and $\mu_3$ represents the constraints in the low-$k$ bin.}
\label{Fig:Constraints_1MG}
\end{figure}

Both traditional and hybrid methods point out some tension in $\Sigma_2$ (2.4-$\sigma$ for traditional binning and 2.2-$\sigma$ for hybrid binning) and $\Sigma_4$ (2.0-$\sigma$ and 2.1-$\sigma$, respectively) when we analyze the 1-D marginalized constraints. These parameters correspond to low and high redshift at small scales, respectively. We find that adding the combination CMBL+DES to Planck+SBR reduces the tension in $\Sigma_2$ to about 1.5-$\sigma$, but the tension in $\Sigma_4$ persists, now at around 2.3-$\sigma$. We plot in Fig. \ref{Fig:Constraints_2MG} the corresponding confidence contours for the growth and lensing MG parameters in each bin using the traditional binning, which leads to tighter constraints compared to the hybrid binning. We can observe that the tension in $\Sigma_2$ is diluted when we add gravitational lensing data, which shifts the contours towards the GR limit. However, as we improve the constraining power, we notice that Planck+SBR+CMBL+DES prefers a value of $\Sigma_4$ higher than 1 with a $\sim$2-$\sigma$ confidence that moves the contours away from GR in the high-$z$, high-$k$ bin. 

Now, we examine the correlations between the MG parameters by calculating the average absolute correlation values in different scenarios. We notice that both traditional and hybrid methods lead to similar correlation values as expected, since the difference between both methods is only a smoothing effect in the transition between bins. Then, using dataset combinations such as Planck+BAO/RSD, Planck+SBR and Planck+SBR+CMBL+DES we find that $\langle|\rho_{\mu_i,\Sigma_j}|\rangle < 0.2$ for the correlation between $\mu_i$ and $\Sigma_j$ parameters, where $\rho_{\mu_i,\Sigma_j}$ represents the Pearson correlation coefficient between the parameters $\mu_i$ and $\Sigma_j$. Additionally, looking at the correlations between the same type of MG parameters, we obtain that $\langle|\rho_{\mu_i,\mu_j}|\rangle, \langle|\rho_{\Sigma_i,\Sigma_j}|\rangle \sim 0.3$ or less. Hence, correlations between the parameters $\mu_i$ and $\Sigma_j$ are not that significant, on average, while the correlations between the MG growth-growth and MG light-light type parameters can become significant even though they may correspond to different bins. 

In order to lower the extra degrees of freedom to produce tighter constraints and further explore MG parameter tensions, we proceed to set some of the MG parameters to their GR values of 1, as was done in \cite{CGQ_MG2020}. We justify this procedure in the following way: Although it was argued in \cite{Huang_2019} that cosmological analyses should study the correlation between different extra parameters in the model instead of varying them one at the time (since correlations might not be negligible), in our case we find that while MG growth-growth and MG light-light type parameters have some moderate degree of correlation, overall it is not the case for the correlation between the $\mu_i$ and $\Sigma_j$ parameters. Thus, we choose to set each of the two parameters equal to 1, one at the time. 

Hence we focus on the rest of the models in Table \ref{Table:Models} with the characteristic that some MG parameters are set to be equal to their GR value of 1. It is important to mention that as we reduce the number of degrees of freedom, we get tighter constraints since the MG parameters follow a one-to-one relationship as explained in \cite{CGQ_MG2020}. So, the constraints on the MG parameter sensitive to a given cosmological probe directly affects the other MG parameter. Additionally, we obtain a better convergence for the chains since the number of extra free parameters is reduced. We show the MCMC statistical results for models \textbf{P2}, \textbf{P3} and \textbf{P4} in Table \ref{Table:Constraints_1MG} and plot the corresponding confidence contours for each bin in Fig. \ref{Fig:Constraints_1MG}. Model \textbf{P2} is subjected to $\eta=1$, so the remaining MG parameters will have $\mu=\Sigma$, so a given cosmological probe that constrains $\Sigma$ should also affect $\mu$. For this model we find that both traditional and hybrid binning methods show some mild tensions above 1-$\sigma$ in high-$k$, low-$z$ and high-$k$, high-$z$ bins. In particular, we observe that the traditional binning method indicates a tension above 2-$\sigma$ in $\mu_2=\Sigma_2=1.14^{+0.12}_{-0.13}$ for Planck+SBR. If we add lensing data we find that the tension in high-$k$, low-$z$ bin is alleviated but the weak tension of 1.5-$\sigma$ is unaffected  in the high-$k$, high-$z$ bin. These findings are similar to those of model \textbf{P1}, where lensing data can alleviate the tension in $\Sigma_2$ but not in $\Sigma_4$. Opposite to this behaviour, model \textbf{P3}, which satisfies $\Sigma=1$, indicates a good agreement with GR with only some mild tensions in $\mu_2$ and $\mu_4$ close to 1-$\sigma$ for the combination Planck+SBR+DES. 

Finally, model \textbf{P4} with $\mu=1$ is sensitive to modifications in the behaviour of the bending of light and a slip effect between the two gravitational potentials while assuming the growth rate of structure is predicted by GR. As expected, adding SNe+BAO to Planck+BAO/RSD does not change the constraints in the $\Sigma_i$ parameters since these are mostly constrained by gravitational lensing data. However, model \textbf{P4}, for both traditional and hybrid binning, indicates a tension of around 2.1-$\sigma$ for $\Sigma_2$ and $\Sigma_4$ (with a maximal 2.3-$\sigma$ tension for $\Sigma_2$ using traditional binning) when we constrain the parameters with Planck+SBR. The addition of CMBL to Planck+SBR lowers the tension in $\Sigma_4$ to 1.1-$\sigma$ while the tension in $\Sigma_2$ goes to 1.9-$\sigma$ for traditional binning (or 1.4-$\sigma$ for hybrid binning). On the other hand, Planck+SBR+DES removes the tension in $\Sigma_2$ but the tension in $\Sigma_4$ increases to 2.2-$\sigma$ with hybrid binning. Using Planck+SBR+DES with traditional binning also removes the tension in $\Sigma_2$ and leads to a tension of 1.9-$\sigma$ in $\Sigma_4$. Therefore, the full dataset combination Planck+SBR+CMBL+DES shows no tension above 2-$\sigma$ in $\Sigma_2$ but reveals a tension in $\Sigma_4$ between 1.9-$\sigma$ and 2.0-$\sigma$ depending on the binning method used after marginalization.

\subsection{Comparison of results from MG functional form parameterization and binning methods}\label{Sec:Reconstruction}
It has been shown that binning over the MG parameters in redshift can reconstruct the constraints using power-law scale-independent parameterizations for $\mu$ and $\eta$ \cite{2019ApJ...871..196L}. Here we are interested in comparing how a scale-independent functional form and a scale-dependent functional parameterization fit the data compared with the binning methods. For the scale-independent functional form, we only consider the time-dependent proportionality to the dark energy density parameter $\propto \Omega_{\text{DE}}(a)$, while (\ref{muEvolution}) and (\ref{SigmaEvolution}) describe the evolution of the MG parameters when scale-dependence is assumed. We compare the constraints individually for $\mu(a,k)$ and $\Sigma(a,k)$ using the models \textbf{P3} and \textbf{P4}, respectively, so we obtain better constraints on each of the parameters individually.

Now, since the MG parameters are a function of time and scale in their most general form, we perform a comparison by analyzing two different scenarios: Firstly, we vary the MG parameters in redshift for fixed values of $k$ corresponding to small and large scale regimes. Secondly, we study the behaviour of the MG parameters in scale for given redshift values. It is worth noticing that the binning methods provide the best fit for the MG parameters per bin, where each bin has a length in redshift and is sensitive to either small scales or large scales. However, the functional form must constrain the MG parameters for all redshift values.

\begin{figure}[t]
\centering
\begin{tabular}{c c}   {\includegraphics[width=7.31cm]{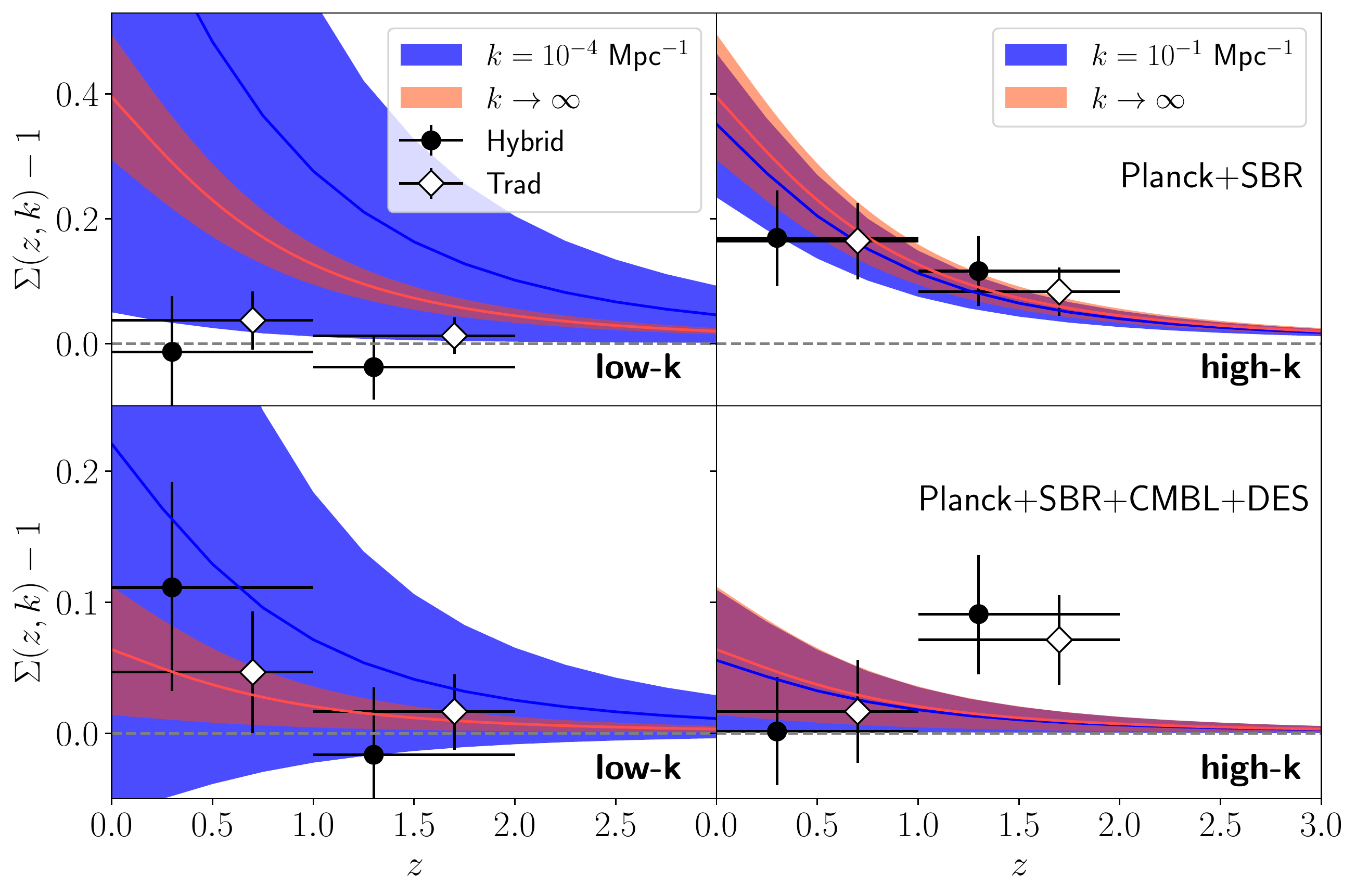}} &
  {\includegraphics[width=7.31cm]{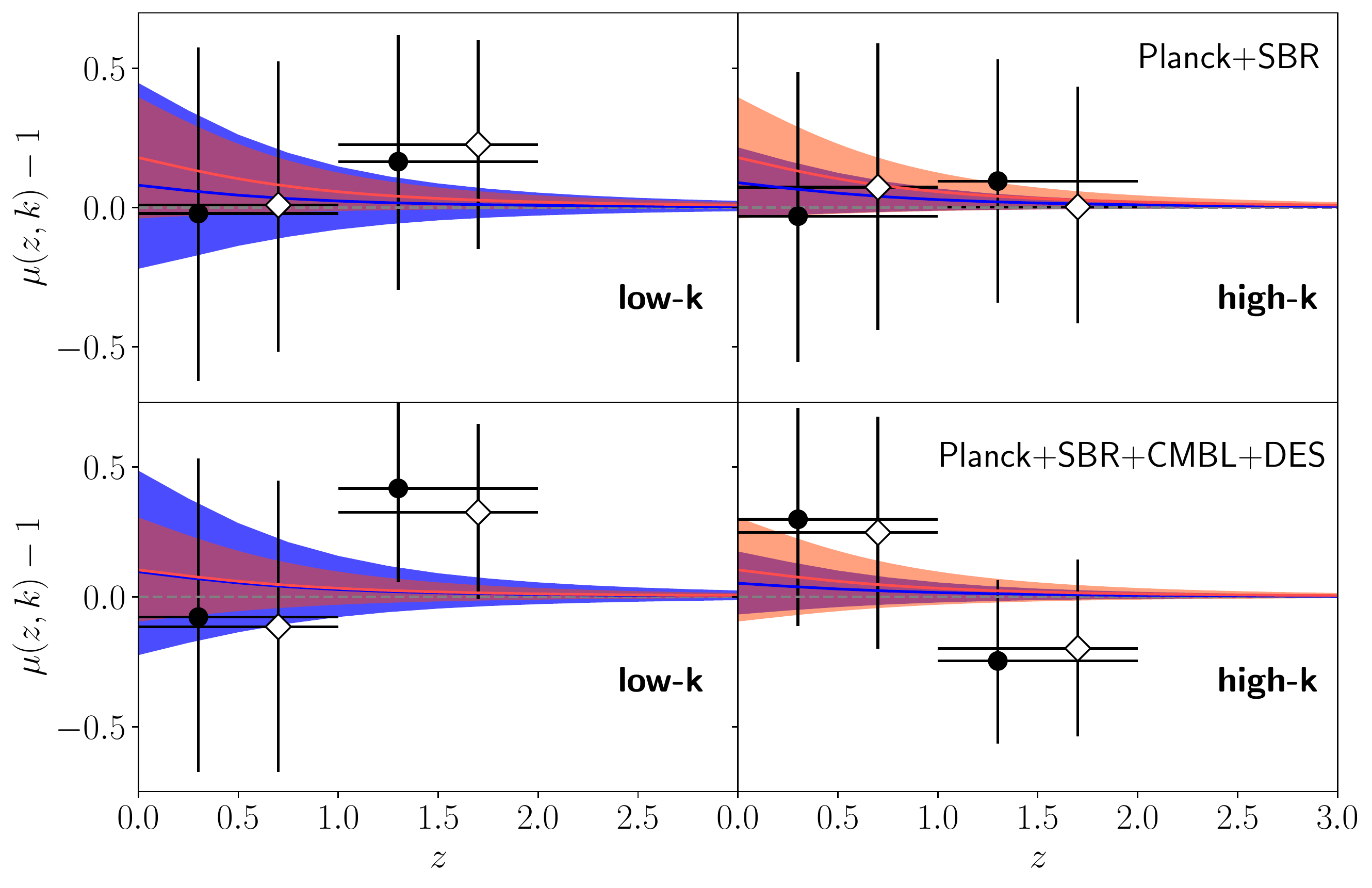}} \\ 
\end{tabular}
\caption{Reconstruction of the functional form in the redshift given in \ref{sec:functional} and comparison with binning methods for the models \textbf{P4} (left figure) and \textbf{P3} (right figure). The constraints shown in the top panels use the Planck+SBR combination while the constraints in the bottom include lensing data as well. We plot the constraints for the low-k regime in the left panels and we show the reconstruction at high-k in the right panels. The figures show the 68\% confidence contours for the functional form with scale dependence (blue band) and no scale dependence (red band). We match the constraints of the functional form with the hybrid (black dots) and traditional (white diamonds) binning methods. The redshift bin width and the mean value of the MG parameter are represented with the horizontal bars while the vertical error bars indicate the 68\% limits. The horizontal gray dashed line indicates the GR value for the MG parameter. We observe that when looking at $\Sigma(z,k)$, among other discrepancies between functional and binning predictions, the addition of lensing data reveals that the binning methods manifest a higher deviation with respect to GR in the high-$z$, high-$k$ bin than what the functional form predicts. However, the functional form and binning methods both agree relatively well when considering constraints on $\mu$, where they both also agree with GR.}
\label{Fig:MG_reconstruction_redshift}
\end{figure}

We show in Fig. \ref{Fig:MG_reconstruction_redshift} the reconstruction of the functional form in redshift and compare it with binning methods. We plot the functional form for the MG parameters with a given scale value, and we choose to represent the low-$k$ region by $k=10^{-4}\text{Mpc}^{-1}$ since it is far enough from the transition scale between bins at $k=10^{-2}\text{Mpc}^{-1}$. Now, for the high-$k$ regime we choose $k=10^{-1}\text{Mpc}^{-1}$ since higher values may represent a region where the linear matter power spectrum and the nonlinear matter power spectrum start to split. We notice that the traditional and hybrid method constraints agree well with each other within 1-$\sigma$ in all the bins. When constraining $\Sigma$ we can observe that the functional form based on $\propto \Omega_{\text{DE}}(a)$ fails to match the binning methods constraints in at least one bin. Indeed, Planck+SBR indicates that the functional form reconstructs exactly what the binning methods predict at small scales, while at large scales the functional form $\propto \Omega_{\text{DE}}(a)$ predicts a higher tension with respect to GR. Allowing a scale-dependent factor multiplied next to $\Omega_{\text{DE}}(a)$ leads to an agreement of both functional form and binning methods at all redshift bins when using Planck+SBR. A similar situation occurs when considering Planck+SBR+CMBL+DES at low-$k$. However, we observe that the functional form with or without scale dependence fails to match the constraints of the binning methods within 1-$\sigma$ in one bin. More precisely, we find a disagreement in the high-$k$, high-$z$ bin when using Planck+SBR+CMBL+DES, where the binning methods indicate a higher tension with respect to GR than the functional form. On the other hand, the functional form seems to agree well with both binning methods when constraining $\mu$ in redshift.

\begin{figure}[t]
\centering
\begin{tabular}{c c}   {\includegraphics[width=7.31cm]{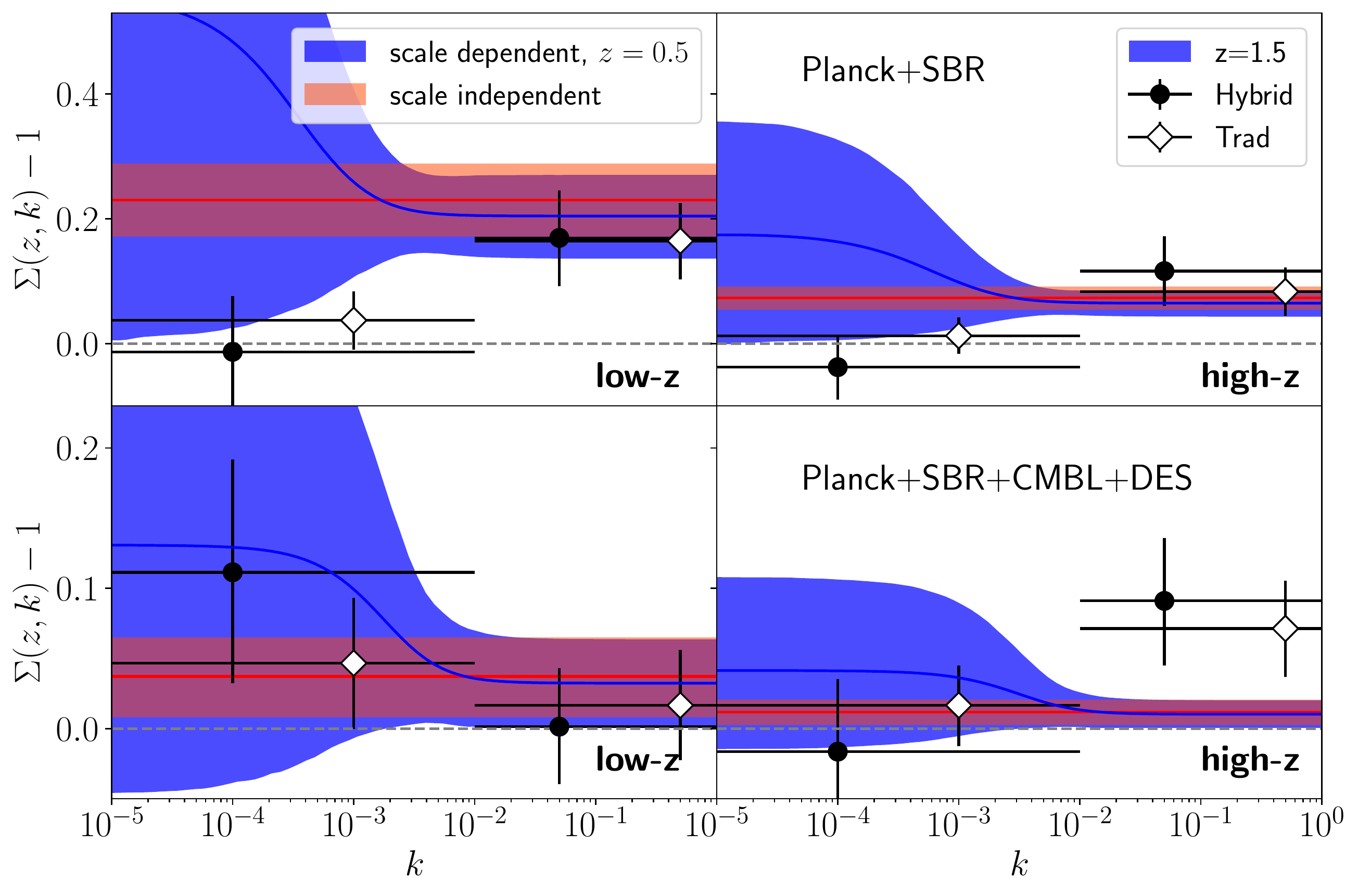}} &
  {\includegraphics[width=7.31cm]{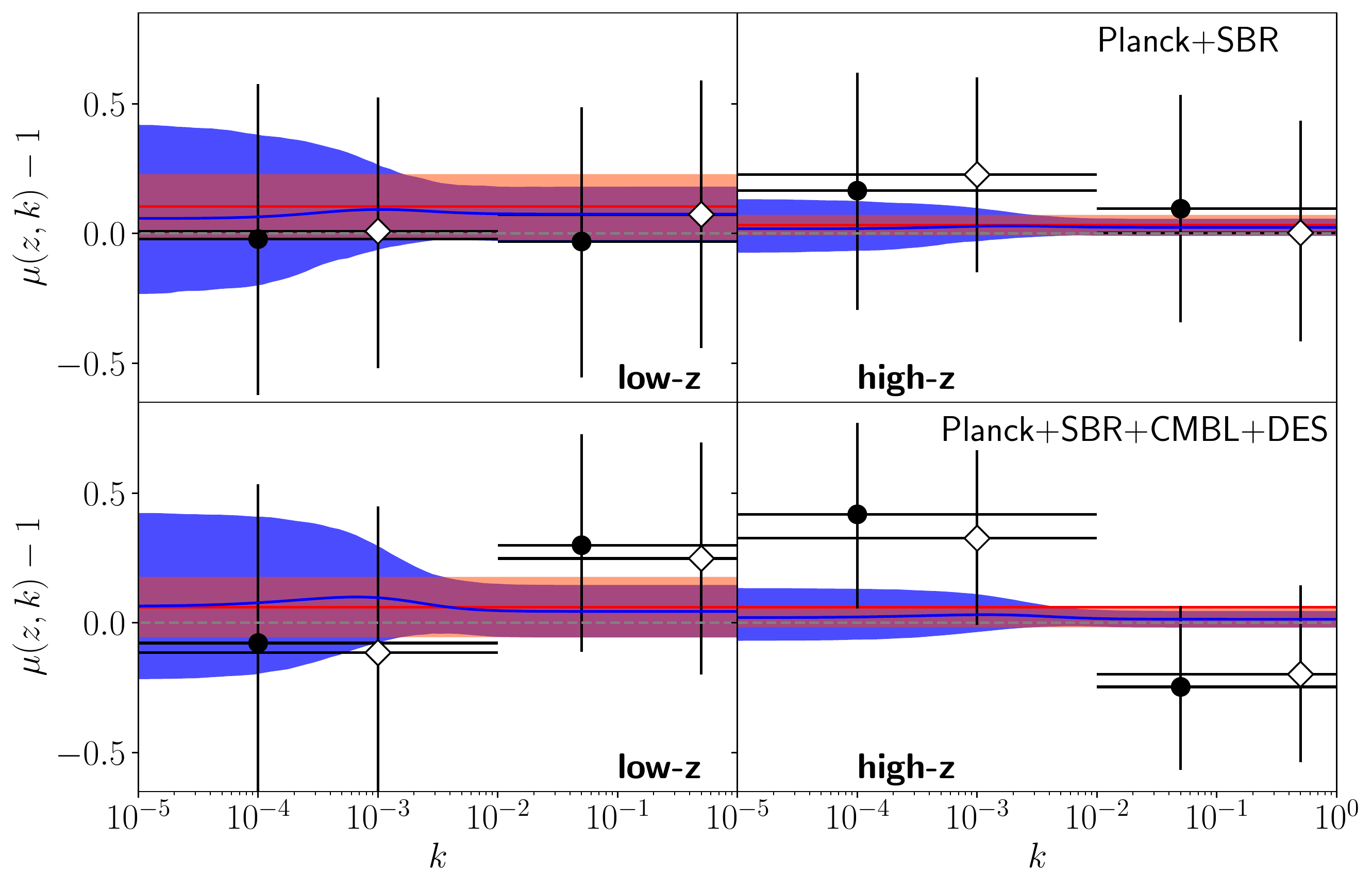}} \\ 
\end{tabular}
\caption{Analogous reconstruction of Fig. \ref{Fig:MG_reconstruction_redshift} for scale instead of redshift. The left figure corresponds to model \textbf{P4} while the figure on the right shows the constraints for model \textbf{P3}. The 68\% confidence regions of the functional form using binning methods are shown. The red band indicates constant values meaning that the MG parameter does not depend on the scale, while the blue band shows the constraints when allowing scale-dependence on the MG parameter. The black dots show the constraints of the hybrid binning method and the white diamonds the constraints using the traditional binning method. The 68\% limits of binning methods are represented with vertical error bars, and the scale bin widths are represented with horizontal bars. Here, the functional form cannot match the binning methods for $\Sigma(z,k)$ in the high-$z$, high-$k$ bin, a region that is constrained exclusively by the $\Omega_{\text{DE}}$ parameter. On the other hand, the binning and functional forms are all in agreement with each other and with GR when constraining $\mu$.}
\label{Fig:MG_reconstruction_scale}
\end{figure}

We now focus on a similar comparison of functional form and binning methods, but as a function of the scale. We show in Fig. \ref{Fig:MG_reconstruction_scale} a plot analogous to the one presented in Fig. \ref{Fig:MG_reconstruction_redshift}, now varying the MG parameters in scale within the redshift bins. Here we vary the MG parameters for the functional form setting $z=0.5$ and $z=1.5$, which are the mean redshift values for the low-$z$ and high-$z$ bins, respectively. Since the functional form based only on $\propto \Omega_{\text{DE}}(a)$ is not a function of scale, the 68\% confidence limits are represented as horizontal bands over the scale. For Planck+SBR we find that a parameterization exclusively based on $\propto \Omega_{\text{DE}}(a)$ does not match the predictions from binning methods at large scales when constraining $\Sigma$. Using Planck+SBR+CMBL+DES shows that all methods agree at low-$z$, but a discrepancy between the functional form and the binning methods occurs in the high-$z$ bin for small scales, which shows a tension at about 2-$\sigma$ if we use binning methods while the functional form indicates an agreement with GR in that bin. It is worth noticing that this disagreement between the functional form and the binning methods in the high-$z$, high-$k$ bin is due exclusively to the $\Omega_{\text{DE}}$ proportionality assumed for the MG parameters, since the scale dependence only affects the MG parameters at low-$k$. Therefore, using a different redshift dependence for the functional form may lead to a better agreement with the binning methods. On the other hand, we observe that the binning methods and the functional form agree when constraining $\mu$, and both seem to be in concordance with GR.

\subsection{Model comparison}
\begin{figure}[t]
\centering
\begin{tabular}{c c}   {\includegraphics[width=7.4cm]{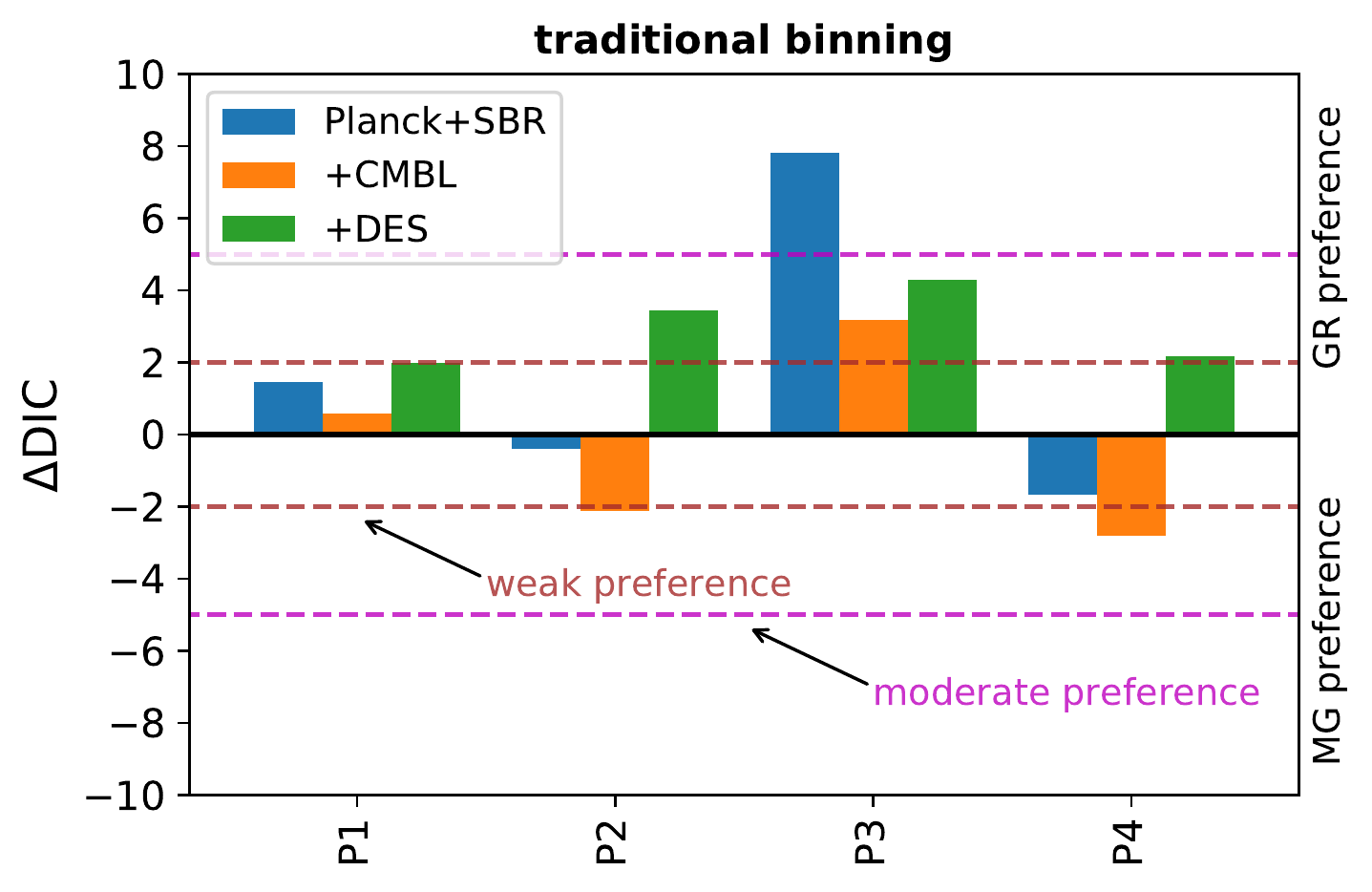}} &
  {\includegraphics[width=7.4cm]{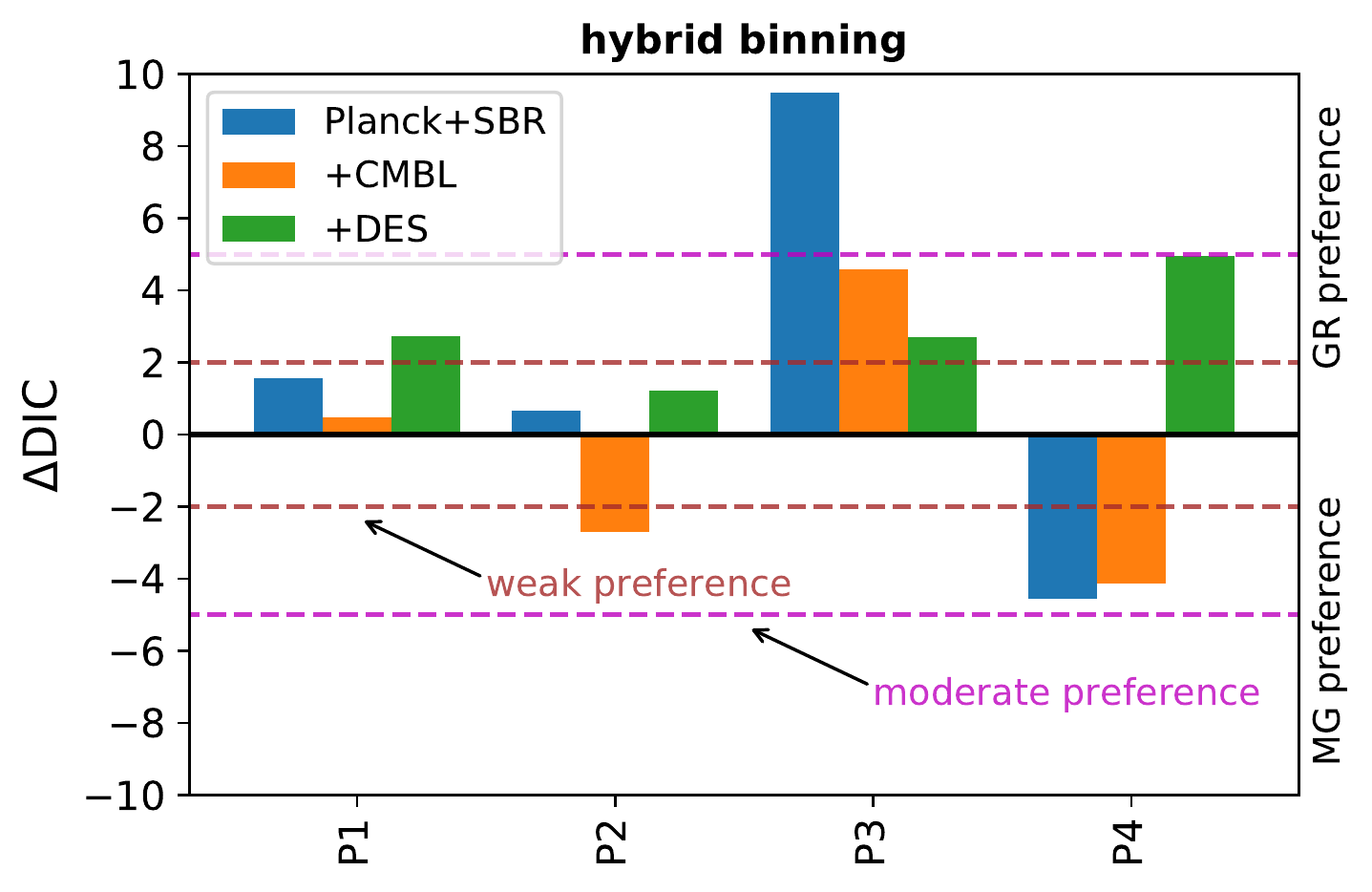}} \\ 
\end{tabular}
\caption{Values of $\Delta \text{DIC} = \text{DIC}_{\text{MG}} - \text{DIC}_{\Lambda\text{CDM}}$ given the models listed in Table \ref{Table:Models}. The color bars represent the values of $\Delta\text{DIC}$ given certain datasets. We adopt the Jeffrey's scale to interpret $\Delta \text{DIC}$ as presented in \cite{2016-tensions-Grandis-etal}, with a modified nomenclature. We interpret $-2<\Delta\text{DIC}<2$ as no significant preference for either model, the range $-5<\Delta\text{DIC}<-2$ represents a weak preference for the MG model (this threshold is represented by a dashed horizontal gray line), $-10<\Delta\text{DIC}<-5$ is a moderate preference (dashed horizontal purple line), while $\Delta\text{DIC}<-10$ indicates a strong preference for the MG model. Analogous values for positive $\Delta\text{DIC}$ are interpreted as a preference for $\Lambda$CDM. Here, only model \textbf{P4} seems to be somewhat favored over $\Lambda$CDM (except when using DES) with $\Delta$DIC values in the weak preference range for MG and close to the moderate preference threshold. On the other hand, $\Lambda$CDM is favoured over \textbf{P3}.}
\label{Fig:DIC}
\end{figure}

Even though tensions with respect to GR can be found for an extended model, it is important to perform a model comparison in order to assess whether or not the extended model is supported by the data with respect to the standard $\Lambda$CDM model. Among the tools for model selection is the Bayesian Evidence, which is commonly used. It is well known that the Bayesian Evidence has an Occam's razor component in the sense that it penalizes models with extra parameters and also depends on the prior volumes. On the other hand, different information criteria have been widely studied, such as the Akaike Information Criteria (AIC), the Bayesian Information Criteria (BIC) and the  Deviance Information Criteria (DIC). Among them, AIC and BIC works with the maximum likelihood value while DIC adds a Bayesian complexity term to the analysis. Due to the prior dependence of the Bayesian Evidence, we opt to use the DIC as a measure for model selection. The DIC was developed as a statistical tool in \cite{DIC-Spiegelhalter,DIC-Spiegelhalter2} and later introduced in astrophysics in \cite{Trotta2008-model-select}. Since then, several works in cosmology have adopted this measure in order to perform a model comparison \cite{2007-Liddle-IC-for-astroph,Martinelli-Hogg,DiValentino-omegak}. The DIC is defined as
\begin{equation}
\text{DIC} = \chi^2_{\text{eff}}(\hat{\theta}) + 2P_D.
\label{DIC}
\end{equation}
Here, $\hat{\theta}$ represents the best fit parameter vector and therefore $\chi^2_{\text{eff}}(\hat{\theta})=-2\ln\mathcal{L}(\hat{\theta})$. The Bayesian complexity is involved in the analysis by $P_D=\overline{\chi^2_{\text{eff}}(\theta)}-\chi^2_{\text{eff}}(\hat{\theta})$, which is effectively the number of parameters of the model for very well-constrained parameters. What ultimately matters from a model comparison perspective is the difference in the DIC between the extended model and the standard model. Therefore, we compute $\Delta \text{DIC} = \text{DIC}_{\text{MG}} - \text{DIC}_{\Lambda\text{CDM}}$, so a negative $\Delta\text{DIC}$ supports MG while a positive value of $\Delta\text{DIC}$ favors GR for a given dataset. 

We introduce the scale we will follow for the interpretation of $\Delta\text{DIC}$ and show the results for some of these values in Fig. \ref{Fig:DIC}. Let us focus on the model \textbf{P1} which has more MG parameters: We find that there is no significant preference for MG over $\Lambda$CDM when we use Planck+SBR or when we combine this dataset with CMBL or DES. In general, the $\Delta$DIC values obtained using the traditional and hybrid methods were similar. A similar situation occurs for model \textbf{P2} where we find no significant preference over $\Lambda$CDM. However, we find that model \textbf{P3} provides higher values of $\Delta$DIC and that Planck+SBR has a moderate preference for $\Lambda$CDM over \textbf{P3}. Finally, we observe that there was a weak and near to moderate preference for \textbf{P4} when using Planck+SBR or Planck+SBR+CMBL with $\Delta$DIC$< -4$. However, the measure turns to $\Delta$DIC$\approx 5$ when we consider Planck+SBR+DES.

Finally, it is worth mentioning that the factor $\chi^2_{\text{eff}}(\hat{\theta})$ accounts for the goodness of fit while $P_D$ penalizes models with extra parameters. Therefore, we expect that adding more bins (and then more parameters) to the extended models in Table \ref{Table:Models} consequently disfavors the extended models with respect to $\Lambda$CDM. However, we point out that the goal of the binning methods is to provide constraints on the MG parameters in a model-independent fashion, while also giving hints on how the MG parameters should evolve within each redshift (and scale) bin. Hence, a suitable functional form can eventually be adopted in such a way that it matches the binning method predictions with a minimum number of extra parameters. Nevertheless, we notice that model \textbf{P4}, which adds 4 extra parameters, is favored over $\Lambda$CDM with some datasets.

\section{Summary and concluding remarks
\label{sec:conclusion}}

In order to explore MG parameter constraints from current data without limitations that could be associated with a functional form, we performed here an analysis using binning methods in redshift and scale. 

Furthermore, for comparison, we performed an analysis of the functional form for MG parameters which include redshift and scale dependence. For some models, we vary one MG parameter at a time, which allows more constraining power from the current available datasets. 

We reported results for each method as well as their comparisons with each other along with a model selection analysis between MG models and $\Lambda$CDM. In summary, we found the following results:

\begin{itemize}

\item The redshift- and scale-dependent functional form constraints show a moderate tension of about 2.9-$\sigma$ at scales of $k\geq 10^{-2}\text{Mpc}^{-1}$ when using Planck+SBR data. These tensions go away when adding lensing data, i.e. CMBL+DES. 

\item When fixing one MG parameter, the redshift- and scale-dependent functional form constraints were tighter and revealed that while there is no tension for the parameter $\mu$, the MG parameter $\Sigma$ shows tension with a maximal 3.5-$\sigma$ when using Planck+SBR at scales $k\geq 10^{-2}\text{Mpc}^{-1}$. These tensions are diluted as we go to larger scales due to enlarged error bars, or go away when using lensing data, i.e. CMBL+DES. 

\item Moving to the binning methods, when fitting the full set of MG parameters (i.e. before fixing any MG parameters and denoted as model \textbf{P1}), we find that $\Sigma_2$ shows a mild tension and $\Sigma_4$ shows a more persistent tension as follows:
    \begin{itemize}
        \item $\Sigma_2$ shows a 2.2-$\sigma$ (hybrid binning) or a 2.4-$\sigma$ (traditional binning) tension when using Planck+SBR, but this reduces to a mild 1.5-$\sigma$ tension when lensing data is included.
        \item For $\Sigma_4$, we have a 2.0-$\sigma$ (traditional binning) or 2.1-$\sigma$ (hybrid binning) tension when using Planck+SBR, and a 2.3-$\sigma$ tension when using Planck+SBR+CMBL+ DES.
    \end{itemize}

\item Next, we fix one MG parameter and vary the other, with the various models shown in Table \ref{Table:Models}. This reduces the extra number of parameters and provides better constraints, as expected. We find that model \textbf{P3} (where $\Sigma$ is fixed to 1) is in agreement with GR while model \textbf{P2} (i.e. $\eta$ is fixed to 1) and model \textbf{P4} (where $\mu$ is fixed to 1) seem to give overall similar constraints with respect to each other; in particular, \textbf{P4} reproduces the tension in $\Sigma_2$ and $\Sigma_4$ using Planck+SBR, and the addition of lensing data only alleviates tension in $\Sigma_2$ while a mild tension of 1.9-$\sigma$ to 2.0-$\sigma$ persists in $\Sigma_4$ using both binning methods. 

\item Some important disagreements are found when comparing the functional form of the MG parameters versus the binning methods. A reconstruction method and comparison shows that for the parameter $\Sigma$, a scale-independent parameterization based on $\Omega_\text{DE}(a)$ does not match the constraints from the binning methods in all bins. Adding a scale dependence to it leads to a partial agreement in some cases, although we find an irreconcilable disagreement between the two in the high-$z$, high-$k$ bin. This can be attributed to the lack of a suitable parameterization that properly captures deviations from GR in these regimes, as seems to be the case when assuming a MG time evolution $\propto\Omega_\text{DE}(a)$.
    
\item From performing a $\Delta$DIC model comparison analysis, we find that the model \textbf{P4} (i.e. $\mu=1$) is weakly favored over $\Lambda$CDM when DES data is not included. This is similar to a previous result when using the functional and time-dependent-only $\Sigma(a)$ parameter \cite{CGQ_MG2020}, with the exception that now CMBL does not exclude the MG model. We find a weak preference for $\Lambda$CDM over model \textbf{P1}, and there was no clear preference between model \textbf{P2} ($\eta=1$) and $\Lambda$CDM. However, $\Lambda$CDM is clearly favored over \textbf{P3} ($\Sigma=1$).

\end{itemize}

\section*{Acknowledgements}
M.I. acknowledges that this material is based upon work supported in part by the Department of Energy, Office of Science, under Award Number DE-SC0019206. CGQ gratefully acknowledges a PhD scholarship from the Mexican National Council for Science and Technology (CONACYT). ON gratefully acknowledges the Physics REU (Research Experiences for Undergraduates) program at the University of Texas at Dallas, funded by the National Science Foundation (NSF).

The authors acknowledge the Texas Advanced Computing Center (TACC) at The University of Texas at Austin for providing HPC resources that have contributed to the research results reported within this paper. URL: http://www.tacc.utexas.edu.

\bibliographystyle{JHEP}
\bibliography{references}
\end{document}